\documentclass[10pt,a4paper]{article}
\usepackage[english]{babel}
\usepackage[latin1]{inputenc}
\usepackage{amsfonts,amsbsy,bm,euscript,mathrsfs}
\usepackage{amssymb,stmaryrd,faktor}
\usepackage[tbtags]{amsmath}
\usepackage{bbm}
\usepackage{graphicx}
\usepackage[title,titletoc]{appendix}
\usepackage[bookmarks=true,colorlinks=true,linkcolor=blue,citecolor=blue,urlcolor=blue,bookmarksnumbered]{hyperref}

\usepackage{dsfont}
\usepackage{collref}
\usepackage{lmodern}
\usepackage{mathrsfs}
\usepackage{mathtools}
\usepackage{bbm}
\usepackage{braket}
\usepackage{slashed}
\usepackage{float}
\usepackage{graphicx}
\usepackage{booktabs}
\usepackage{subfig}
\usepackage{tikz}
\usepackage[symbol]{footmisc}
\usepackage{hyperref}
\usepackage{cleveref}
\usetikzlibrary{plotmarks,calc,decorations,decorations.pathmorphing}
\usepackage[normalem]{ulem}

\numberwithin{equation}{section}

\makeatletter
\renewcommand\section{\@startsection {section}{1}{\z@}
{-3.5ex \@plus -1ex \@minus -.2ex}
{2.3ex \@plus.2ex}
{\normalfont\Large\bfseries}}
\renewcommand\subsection{\@startsection{subsection}{2}{\z@}
{-3.25ex\@plus -1ex \@minus -.2ex}
{1.5ex \@plus.2ex}
{\normalfont\large\bfseries}}
\makeatother

\newcommand{\arXivlink}[1]{\href{http://arXiv.org/abs/#1}{arXiv:#1}}

\newcommand{\alg}[1]{\mathfrak{#1}}

\usepackage{blkarray}

\begin{document}

\thispagestyle{empty}
\begin{flushright}\footnotesize\ttfamily
DMUS-MP-23/02
\end{flushright}
\vspace{2em}

\begin{center}

{\Large\bf \vspace{0.2cm}
{\color{black} \large A study of integrable form factors in massless relativistic $AdS_2$}} 
\vspace{1.5cm}

\textrm{Daniele Bielli$^{a,b}$, Vaibhav Gautam$^{a}$, Alessandro Torrielli$^{a}$ \footnote[1]{\textit{E-mail addresses:} \texttt{d.bielli@surrey.ac.uk, v.gautam@surrey.ac.uk, a.torrielli@surrey.ac.uk}}}
\vspace{0.8cm}
\\
$^a$ \textit{Department of Mathematics, University of Surrey, Guildford, GU2 7XH, UK}
\\
\vspace{0.3cm}
$^b$ \textit{Dipartimento di Fisica, Universit{\`a} degli studi di Milano--Bicocca,\\ and INFN, Sezione di Milano--Bicocca, Piazza della Scienza 3, 20126 Milano, Italy}




\end{center}

\vspace{2em}

\begin{abstract}\noindent 
In this paper we initiate the study of form factors for the massless scattering of integrable $AdS_2$ superstrings, where the difference-form of the $S$-matrix can be exploited to implement the relativistic form factor bootstrap. The non-standard nature of the $S$-matrix implies that traditional methods do not apply. We use the fact that the massless $AdS_2$ $S$-matrix is a limit of a better-behaved $S$-matrix found by Fendley. We show that the previously conjectured massless $AdS_2$ dressing factor coincides with the limit of the De Martino - Moriconi improved dressing factor for the Fendley $S$-matrix. After finding a method to construct integral representations of relativistic dressing factors satisfying specific assumptions, we obtain analytic proofs of crossing and unitarity relations and propose a solution to the form factors constraints in the two-particle case.
\end{abstract}

\newpage

\overfullrule=0pt
\parskip=2pt
\parindent=12pt
\headheight=0.0in \headsep=0.0in \topmargin=0.0in \oddsidemargin=0in

\vspace{-3cm}
\thispagestyle{empty}
\vspace{-1cm}

\tableofcontents

\setcounter{footnote}{0}

\section{\label{sec:Intro}Introduction}

\subsection{$AdS_2$ integrable scattering}

Let us briefly describe some work that has recently taken place as regards to the $AdS_2 \times S^2 \times T^6$ string theory background \cite{ads2, ads21, ads22, ads23, ads24} for what concerns its integrability description. The duality is in this case particularly mysterious, and it should involve a certain superconformal quantum mechanics or a particular chiral 2D CFT \cite{dual,dual1, dual2, dual3, dual4, gen, gen1, gen2, gen3, gen4, gen5, gen6, gen7, gen9, gen10, gen11, gen12, gen13, gen14, gen15}. The coset action is based on $\frac{PSU(1,1|2)}{ SO(1,1) \times SO(2)}$ \cite{Metsaev:1998it, Metsaev:1998it1, Metsaev:1998it2,gen10}. This has been shown to have classical integrability \cite{BPR} up to a second order expansion in the fermionic fields \cite{Sorokin:2011rr,Cagnazzo:2011at}.

The integrable $S$-matrix for this theory was constructed in \cite{Hoare:2014kma} by postulating a centrally-extended $\mathfrak{psu}(1|1)^2$ symmetry as the symmetry remaining after the choice of a BMN vacuum \cite{Berenstein:2002jq,amsw, amsw1, amsw2, Per6, Per10}, and adapting to it the familiar integrable AdS/CFT Hopf-algebraic construction \cite{BeisReview,beis0, beis01, beis02, beis03, beis04}. The massive $S$-matrix was shown to satisfy the properties of crossing and unitarity, however fixing the appropriate dressing factor for massive particles is still an open problem. In the massive sector there is agreement with the perturbative calculations available in the literature \cite{amsw, amsw1, amsw2, Per6, Per10}. The massive particle representations are of {\it long} type, as opposed to what happens in higher-dimensional $AdS$s. 

In $AdS_2$ only the massless particle representations are {\it short}. The massless $S$-matrix can be derived via a limiting procedure from the massive $S$-matrix \cite{Zamol2, Zamol21}, and it displays the typical right- and left-mover (chirality) splitting. An analysis of the Yangian symmetry, with a primary focus on the massive representations, is presented in \cite{Hoare:2014kma,Hoare:2014kmaa}. The massless sector is less amenable to a matching with perturbation theory \cite{amsw, amsw1, amsw2, Per6, Per10, MI}. It is important to point out that massless scattering is rather different from its massive counterpart \cite{Borsato:2016xns}, and in relativistic models it is traditionally associated with the RG flow between two-dimensional conformal field theories \cite{Zamol2,Zamol21}. 

The BMN limit is trivial for massive particles and also for massless particles of opposite chirality \cite{DiegoBogdanAle,AA1}. The BMN limit is non-trivial between right-right and left-left movers \cite{AA1} and it describes a conformal field theory, where left and right movers are decoupled, with ${\cal{N}}=1$ supersymmetry. However, the S-matrix, for which a dressing factor has been conjectured \cite{AA1} and whose validity extends to the whole massless sector, is not the standard one of ${\cal{N}}=1$ theories \cite{Fendley:1990cy,Schoutens,MC}, although it is of eight-vertex type \cite{Baxter:1972hz, Baxter:1972hz1}. As a consequence of this fact (lack of $U(1)$ symmetry), there is no reference state (pseudovacuum) to setup the algebraic Bethe ansatz \cite{Levkovich-Maslyuk:2016kfv}. The Bethe equations have been conjectured from string theory \cite{Sorokin:2011rr}. There is a vast literature on the problem of Bethe ansaetze without $U(1)$ symmetry \cite{Nepo, Nepo1, Nepo2, Nepo3, Nepo4, Nepo5, Nepo6}. In \cite{AA1} the {\it free-fermion condition} \cite{MC,Ahn:1993qa} and the technique of {\it inversion relations} \cite{Zamolodchikov:1991vh} were exploited to obtain a set of auxiliary Bethe equations, which were then compared with the ones from string theory. In \cite{alea} more evidence was found by brute-force diagonalising the transfer matrix for a maximum of 5 particles, and Yangian symmetry was also discussed for this superconformal scattering theory.  The free-fermion condition was then used again in \cite{Marius} to recast the problem in a different form, using a suitable Bogoliubov transformation. In this form a state was found for the two-particle transfer matrix, effectively playing the role of the pseudovacuum (this state was dubbed pseudo-pseudovacuum in \cite{Marius}).
In \cite{AA2} a change of variable was found which recasts the massless scattering problem in difference form. This applies both to $AdS_3$, where the new variable has been further exploited in \cite{gamma2,AleSSergey}, and $AdS_2$. In \cite{AA2} further properties of the massless $AdS_2$ dressing factor have been determined.

In recent literature \cite{Marius2}, which connects to a series of works \cite{Dublin}, a very interesting $S$-matrix and symmetries have been found which interpolate\footnote{We thank Ben Hoare for discussions on the interpolation procedure.} between $AdS_2$ and $AdS_3$.

\subsection{This paper}

In this paper we initiate the study of form factors \cite{a1,a2,a3,Babu,BabuF,DelfinoMussSimo,Karo} associated with the $AdS_2$ integrable scattering problem. No attempt at this has been made so far to our knowledge, either in the domain of the hexagon approach \cite{Basso:2013vsa,hexagon,Eden:2021xhe}, or using the more traditional approach (cf. \cite{Thomas}), which in the relativistic case is based on Smirnov's axioms (to be reviewed in the next section). We shall focus on the latter, and restrict ourselves to the massless sector, where difference form is restored. Moreover, we focus on the left-left and right-right scattering, which describes a conformal field theory in the BMN limit. By exploiting the difference form we can make progress on the explicit analytic computation of form factors based primarily on the criterion of meromorphicity.

The $S$-matrix is of eight-vertex type, in that it has all the entries which are allowed by fermion-number conservation (modulo 2) different from zero. In bosonic models this would be stated as the absence of $U(1)$ symmetry. A recent account is reported in \cite{Lal:2023dkj}. This is accompanied by the lack of a highest/lowest weight state in the representation. The Bethe equations have to be derived with methods alternative to the algebraic Bethe ansatz, as done in \cite{AA1}. This also means that the task of finding solutions to the relations, which the form factors have to satisfy, is rather complicated. In particular, the off-shell Bethe ansatz method which was developed in \cite{Babu,BabuF} is here not available, due to the absence of a pseudo-vacuum. Supersymmetry complicates the task even further. In the literature similar models have been attacked before, and in particular the two-particle form factors have been the focus of attention \cite{Ahn:1993qa,Mussardo:1998kq}. Our model does not fall in the class that has been studied before. We exploit the fact, observed in \cite{AA1}, that our $S$-matrices are limits of the Fendley $S$-matrix \cite{Fendley:1990cy}, for which the form factors are also not known. We construct the necessary minimal two-particle form factors explicitly in this paper, and argue that they satisfy the appropriate axioms. Nevertheless, we can still employ a method which has been refined by Mussardo \cite{Mussardo:1998kq}, and apply it to our situation. In order to tackle our specific problem we also construct a method for finding implicit, and, in some simple cases, also explicit, integral expressions of relativistic dressing factors. This method turns out to be applicable in a quite ample array of cases - provided that certain hypotheses are satisfied. With a combination of the Mussardo method and the integral representations which we construct, we obtain the minimal two-particle form-factors for the Fendley $S$-matrix with the De Martino - Moriconi improved dressing factor \cite{DM}.

We then show numerically for the first time that the dressing factors postulated in \cite{AA2} coincide with the limits of the Fendley dressing factors, in the improved version by De Martino and Moriconi. The fact that the right-hand side of the crossing and unitarity equations tend to the $AdS_2$ ones was known from \cite{AA1}, but here we show that the dressing factors of \cite{AA1,AA2} solving such conditions are also precisely the asymptotic limits of the De Martino - Moriconi dressing factor.   

Our results open a new direction, which consists of obtaining predictions for the $AdS_2$ form factors as limits of the Fendley form factors, by-passing the limitations intrinsic to massless $AdS_2$. The $AdS_2$ $S$-matrix in the massless case is not braiding unitary, but rather satisfies a combined condition between the so-called solutions 3 and 5. This prevents to apply even the Mussardo method for the form factors, which is the very reason why we have taken a detour via the better-behaved Fendley $S$-matrix. We end the paper with some future directions and an appendix with a few of the mathematical proofs.

\section{Form Factors Recap}

\subsection{Form factors}

Form factors \cite{a1,a2} are an extremely important part of the bootstrap programme, which leads to the complete solution of an integrable quantum field theory. The bootstrap programme starts with determining the exact $S$-matrix and the Hilbert space of quantum states, and then proceeds with the finite-volume analysis and finally establishing the array of all the form factors for the operators in the theory. Using the explicit expressions of the form factors one can then ultimately reconstruct all the $n$-point correlation functions. The formulas which one reaches at the end are then typically tested against the perturbative Feynman-graph expansion - for a review, see \cite{a3,Babu} and references therein. 

In a relativistic setting the $n$-particle form factor, which one associates with a local operator ${\cal{O}}(x,t)$, is defined to be the matrix element
\begin{equation}
\label{in}
F^{\cal{O}}_{\alpha_1 ... \alpha_n} (\theta_1,...,\theta_n) = \langle 0| {\cal{O}}(0)|\theta_1,...,\theta_n\rangle_{\alpha_1 ... \alpha_n}
\end{equation}
(the position is taken to be the origin, and time is suppressed) with $\theta_i$ being the rapidity of the $i$-th particle in the {\it in} state, {\it i.e.} the ket in (\ref{in}), and $\alpha_i$ denote any other internal degree of freedom. We will focus in this paper on a relativistic dispersion relation, with massless right-movers, the internal degree of freedom being a boson {\it vs} fermion label:
\begin{equation}
p_i = E_i = e^{\theta_i}, \qquad \alpha_i \in \{b,f\}, \qquad b = \mbox{boson}, \qquad f = \mbox{fermion}. 
\end{equation}
We have left implicit a mass parameter $M$ in the dispersion, as it will play no role. The scattering theory which we will treat describes a conformal field theory, and the mass or length scale only features when one puts the theory on a finite circle to setup the thermodynamic Bethe ansatz \cite{DiegoBogdanAle}.   

The computation of form factors proceeds axiomatically by requiring that they satisfy a set of constraints, see for instance \cite{Babu} and \cite{BabuF}. The spirit is to by-pass perturbation theory in the first instance, and to directly guess the solution based on these axioms. We list here for convenience those axioms which are relevant to our massless situation.

\begin{itemize}

\item {\it Permutation}
\begin{eqnarray}
&&F^{\cal{O}}_{\alpha_1 ... \alpha_{j-1} \, \beta_{j} \, \beta_{j+1} \, \alpha_{j+2} ... \alpha_n} (\theta_1,...,\theta_{j-1} , \theta_{j} , \theta_{j+1} , \theta_{j+2}, ... \theta_n) = \\
&&\qquad \qquad F^{\cal{O}}_{\alpha_1 ... \alpha_{j-1} \, \alpha_j \, \alpha_{j+1} \, \alpha_{j+2} ... \alpha_n} (\theta_1,...,\theta_{j-1} , \theta_{j+1} , \theta_j, \theta_{j+2}, ... \theta_n) \, S^{\alpha_j \alpha_{j+1}}_{\beta_j \beta_{j+1}} (\theta_j - \theta_{j+1}).\nonumber
\end{eqnarray}
The entries of the $S$-matrix feature in this axiom as
\begin{equation}
S : V_1 \otimes V_2 \longrightarrow V_2 \otimes V_1, \qquad S |v_\alpha (\theta_1) \rangle \otimes |v_\beta (\theta_2)\rangle = S^{\rho \sigma}_{\alpha \beta} (\theta_1 - \theta_2) |v_\rho (\theta_2) \rangle \otimes |v_\sigma (\theta_1)\rangle.
\end{equation}
It is often convenient to define an $R$-matrix associated with the $S$-matrix as
\begin{eqnarray}
S = \Pi \circ R,\qquad R : V_1 \otimes V_2 \longrightarrow V_1 \otimes V_2.
\end{eqnarray}
$\Pi$ being the graded permutation on the quantum states:
\begin{eqnarray}
\Pi |a\rangle \otimes |b\rangle = (-)^{\sigma} \, |b\rangle \otimes |a\rangle.
\end{eqnarray}
The parity $\sigma$ equals $1$ only when both states are fermions, and it equals $0$ in all the other cases.

\item {\it Periodicity}
\begin{eqnarray}
&&F^{\cal{O}}_{\alpha_1 \, \alpha_{2} ... \alpha_{n-1} \, \alpha_n} (\theta_1 + 2i \pi,\theta_{2}, ... \theta_{n-1}, \theta_n) = (-)^\sigma\, F^{\cal{O}}_{\alpha_2 \, \alpha_{3} ... \alpha_{n} \, \alpha_1} (\theta_2 , \theta_{3}, ... \theta_{n} , \theta_1).
\end{eqnarray}
We have denoted by $\sigma$ an appropriate statistical factor which involves the mutual statistics of the operator and the first particle appearing in the {\it in} state.  

\item {\it Lorentz boost}
\begin{eqnarray}
F^{\cal{O}}_{\alpha_1 \, \alpha_{2} ... \alpha_{n-1} \, \alpha_n} (\theta_1 + \Lambda,\theta_{2}+\Lambda, ... \theta_{n-1}+\Lambda, \theta_n+\Lambda) = e^{s \Lambda}F^{\cal{O}}_{\alpha_1 \, \alpha_{2} ... \alpha_{n-1} \, \alpha_n} (\theta_1,\theta_{2}, ... \theta_{n-1}, \theta_n). 
\end{eqnarray}
We have denoted by $s$ the Lorentz spin of the operator $\cal{O}$.

\item {\it Residue at the kinematical singularities}

The idea is always to complexify and utilise the power of analyticity. The form factors have to be meromorphic functions in all of the complex variables $\theta_i$, and may admit a series of poles. They have bound state poles, which do not concern us in the massless case (see also \cite{DelfinoMussSimo}), and kinematical poles. All these poles are simple. The kinematical poles generate the additional constraint 
\begin{eqnarray}
\label{reso}
&&-\frac{i}{2} \, \mbox{Res}_{\theta_1 = \theta_2+i\pi} F^{\cal{O}}_{\bar{\alpha}_2 \, \alpha_{2} ... \alpha_{n-1} \, \alpha_n} (\theta_1,\theta_{2}, ... \theta_{n-1}, \theta_n) = \\
&&\qquad \qquad {\bf C}_{\bar{\alpha}_2\beta_2} \, \Big[\mathbbmss{1} - (-)^\sigma S_{\alpha_n \rho_{n-3}}^{\beta_n \beta_2}(\theta_2 - \theta_n)...S^{\beta_3\rho_1}_{\alpha_3\alpha_2}(\theta_2-\theta_3)\Bigg]F^{\cal{O}}_{\beta_3 \, \beta_{4} ... \beta_{n-1} \, \beta_n} (\theta_3,... \theta_n). \nonumber
\end{eqnarray}
Here, $\bar{\alpha}$ is used to indicate the anti-particle of $\alpha$ and ${\bf C}_{\alpha \beta}$ represents the charge-conjugation matrix. One has again to allow for a statistical factor $(-)^\sigma$ in (\ref{reso}), depending on the mutual statistics of the operator and the first scattering particle. 

\end{itemize}

See \cite{Ale} for a discussion on the conformal case. Some essential references can be found in \cite{a2,Karo}.

The form factor programme has been at the centre of great progress in $AdS/CFT$, starting with $AdS_5/CFT_4$, where the complications of the model have led to the creation of a completely new and tailored hexagon approach \cite{Basso:2013vsa} - see also \cite{hexagon}. The hexagon approach has been extended to $AdS_3$ \cite{Eden:2021xhe}. The more traditional approach, not based on the hexagon but rather on the relativistic form-factor axioms which we have discussed above, was instead originally followed in \cite{Thomas}, where it had to be generalised to the non-relativistic setting. In \cite{Ale} the traditional relativistic bootstrap programme was applied to massless $AdS_3$, where difference form is restored thanks to the change of variable introduced in \cite{AA2} - see also \cite{gamma2,AleSSergey} - and one can proceed to conjecture exact solutions. In this paper we attack the $AdS_2/CFT_1$ case, which also displays difference form via the same change of variables \cite{AA2} and is therefore amenable to the more traditional axiomatic approach. 

\section{Fendley $S$-matrix And Two-Particle Form Factors}

In order to gain some understanding of the $AdS_2$ case it is convenient to first look at a more regular case, although slightly more complicated. This is the Fendley $S$-matrix \cite{Fendley:1990cy} - this $S$-matrix is much more nicely-behaved than the $AdS_2$ one, and it tends to the various versions of the massless $AdS_2$ $S$-matrix in precise limits. It can be considered therefore as a sort of parent model from which we can learn a number of lessons.

Let us focus here on the Fendley $p=\frac{1}{2}$ $S$-matrix, by displaying the associated $R$-matrix:
\begin{eqnarray}
&&R =\Phi_F(\theta) \Big[E_{11} \otimes E_{11} - E_{22} \otimes E_{22} - \tanh\theta (E_{11} \otimes E_{22}-E_{22} \otimes E_{11}) +\nonumber\\
&&\mbox{sech} \theta \, \cosh \frac{\theta}{2} (-E_{12} \otimes E_{21}+E_{21} \otimes E_{12})-\mbox{sech} \theta \, \sinh \frac{\theta}{2} (-E_{12} \otimes E_{12}+E_{21} \otimes E_{21})\Big].
\end{eqnarray}
The matrices $E_{ij}$ are $2\times 2$ matrices with all zeroes, except $1$ in row $i$, column $j$. They form a basis of $2\times 2$ matrices and act on the states $|\phi\rangle \equiv |b\rangle \equiv |1\rangle$ (bosonic excitation) and $|\psi\rangle \equiv |f\rangle \equiv |2\rangle$
(fermionic excitation) as
\begin{eqnarray}
E_{ij}|k\rangle = \delta_{jk}|i\rangle.
\end{eqnarray}
We have also set $\theta \equiv \theta_1-\theta_2$. $\Phi_F(\theta)$ is the dressing factor, satisfying the crossing equation
\begin{eqnarray}
\Phi_F(\theta)\Phi_F(\theta + i \pi) = \frac{1}{1+\frac{\sinh^2 \frac{\theta}{2}}{\cosh^2 \theta}}.
\end{eqnarray}
This descends from imposing crossing symmetry:
\begin{eqnarray}
\Big[C^{-1}\otimes \mathbbmss{1}\Big]R^{str_1}(\theta+i\pi)\Big[C\otimes \mathbbmss{1}\Big] R(\theta) = \mathbbmss{1}\otimes \mathbbmss{1},
\end{eqnarray}
where ${}^{str_1}$ denote supertransposition in the first factor of the tensor product, and $C$ is the charge conjugation matrix:
\begin{eqnarray}
C=\begin{pmatrix}i&0\\0&1\end{pmatrix}.
\end{eqnarray}
An explicit solution is given by \cite{Fendley:1990cy}
\begin{eqnarray}
\label{Fe}
\Phi_F(\theta) &=& 4\bigg[\frac{1}{2} - \frac{\theta}{\pi i }\bigg]^2 \, \prod_{j=1}^\infty \frac{\Big(j-\frac{1}{2}\Big) \, \prod_{k=1}^{3} \Big(3j + \frac{1}{2} - k\Big)}{\Big(2j - \frac{1}{2}\Big)^2 \Big(2j + \frac{1}{2}\Big)^2} \Big(4 j^2 - \big[\frac{1}{2} - \frac{\theta}{\pi i }\big]^2\Big)^2\nonumber\\
&&\times \frac{\Gamma\Big(3j-\frac{5}{2}+\frac{3}{2} \frac{\theta}{\pi i}\Big)\Gamma\Big(3j-1-\frac{3}{2} \frac{\theta}{\pi i}\Big)}{\Gamma\Big(3j-1+\frac{3}{2} \frac{\theta}{\pi i}\Big)\Gamma\Big(3j+\frac{1}{2}-\frac{3}{2} \frac{\theta}{\pi i}\Big)} \frac{\Gamma\Big(j-\frac{1}{2}+\frac{\theta}{2\pi i}\Big)\Gamma\Big(j-\frac{\theta}{2\pi i}\Big)}{\Gamma\Big(j+\frac{1}{2}-\frac{\theta}{2\pi i}\Big)\Gamma\Big(j+\frac{\theta}{2\pi i}\Big)} \ .
\end{eqnarray}
This dressing factor also satisfies quite remarkably
\begin{eqnarray}
\Phi_F(-\theta)\Phi_F(\theta) = \frac{1}{1+\frac{ \sinh^2\frac{\theta}{2}}{\cosh^2 \theta}},\label{bothi}
\end{eqnarray}
which ensures both the braiding and the physical unitarity of the Fendley $R$-matrix:
\begin{eqnarray}
R_{21}(-\theta)R(\theta) = \mathbbmss{1} \otimes \mathbbmss{1}
\end{eqnarray}\
and
\begin{eqnarray}
R^\dagger(\theta)R(\theta) = \mathbbmss{1} \otimes \mathbbmss{1} \qquad \theta \in \mathbbmss{R},\label{ecci}
\end{eqnarray}
respectively, where $R_{21}$ is defined as
\begin{eqnarray}
R_{21}(\theta)  \equiv \Pi \bigl( R(\theta) \bigr),    
\end{eqnarray}
and $\Pi$ is the graded permutation applied to $R$. $\Pi$ acts on the operator $R$ by means of 
\begin{eqnarray}
\Pi(E_{ij} \otimes E_{kl}) = (-1)^{[deg(i)+deg(j)][deg(k)+deg(l)]} E_{kl}\otimes E_{ij},
\end{eqnarray}
$deg(i)$ denoting the fermionic number of the index $i$.
Property (\ref{ecci}) is indeed guaranteed by
\begin{eqnarray}
\Phi_{F}(\theta)\Phi_{F}^*(\theta)=\frac{1}{1+\frac{ \sinh^2\frac{\theta}{2}}{\cosh^2 \theta}}, \qquad \theta \in \mathbbmss{R},
\end{eqnarray}
which reduces to (\ref{bothi}) upon inspection of the explicit formula for $\Phi_F$.
 
It is clear that in the large (positive and negative) rapidity limit one recovers the relativistic $AdS_2$ crossing equation:
\begin{eqnarray}
\Phi_F(\theta)\Phi_F(\theta + i \pi) \to \frac{e^{\pm \frac{\theta}{2}}}{2\cosh \frac{\theta}{2}}, \qquad \theta \to \pm \infty.\label{limo}
\end{eqnarray}

An improved phase was found by De Martino - Moriconi, formulas (3.6) - (3.8) of  \cite{DM}: 
\begin{eqnarray}
&&\Phi_{DM}(\theta) = -2 \sinh^2 \Big( \frac{i\pi}{4}+\frac{\theta}{2}\Big) \, \prod_{j=0}^\infty \frac{\Gamma(\frac{3}{2}+j) \Gamma(\frac{1}{2}+j-\frac{i\theta}{2\pi})\Gamma(1+j+\frac{i\theta}{2\pi})}{\Gamma(\frac{1}{2}+j) \Gamma(\frac{3}{2}+j+\frac{i\theta}{2\pi})\Gamma(1+j-\frac{i\theta}{2\pi})}\nonumber\\
&&\qquad \qquad \qquad \times \frac{\Gamma(\frac{1}{2}+ 2 q j + 2q)\Gamma(\frac{1}{2}+2q j - \frac{iq\theta}{\pi})\Gamma(\frac{1}{2}+2q j +q+ \frac{iq\theta}{\pi})}{\Gamma(\frac{1}{2}+ 2 q j)\Gamma(\frac{1}{2}+2q j +2 q+ \frac{iq\theta}{\pi})\Gamma(\frac{1}{2}+2q j +q- \frac{iq\theta}{\pi})},\label{DMM}
\end{eqnarray}
where
\begin{eqnarray}
q=1+|p|=\frac{3}{2},\label{eccola}
\end{eqnarray}
with an integral representation given by
\begin{eqnarray}
\Phi_{DM}(\theta) = \exp \Bigg[-\int_0^\infty \frac{dt}{t} \, h(t) \, \frac{\sin t \theta \, \sin t(i\pi-\theta)}{\cosh \pi t}\Bigg], \qquad h(t)=\frac{2}{\sinh \pi t} - \frac{1}{\sinh 2\pi t} - \frac{1}{\sinh \frac{\pi t}{q}}. \label{inteo}
\end{eqnarray}
In the \Cref{app:Appendix A} we prove that
\begin{eqnarray}
\Phi_{DM}(\theta)\Phi_{DM}(\theta+i\pi)=\Phi_{DM}(-\theta)\Phi_{DM}(\theta) = \frac{1}{1+\frac{ \sinh^2\frac{\theta}{2}}{\cosh^2 \theta}}\label{prove1}
\end{eqnarray}
and
\begin{eqnarray}
\Phi_{DM}(\theta)\Phi_{DM}^*(\theta)=\frac{1}{1+\frac{ \sinh^2\frac{\theta}{2}}{\cosh^2 \theta}}, \qquad \theta \in \mathbbmss{R}.\label{prove2}
\end{eqnarray}
It is clear that the crossing equation is the same as for $\Phi_F$, hence one recovers $AdS_2$ in the limit (\ref{limo}).

The function $\Phi_{DM}(\theta)$ is meromorphic with poles and zeroes lying on the imaginary axis given as follows:
\begin{itemize}
    \item {\it Poles}
        
        Poles of the dressing factor $\Phi_{DM}$ are given by the poles of gamma functions in the numerator. 
        \begin{align}
            &\theta = -(2m-1)\pi i , \qquad m=1,2,3,\ldots \qquad 2m^{th} \text{ order pole} \nonumber \\
            &\theta = 2m\pi i , \qquad m=1,2,3,\ldots \qquad 2m^{th} \text{ order pole} \nonumber \\ 
            &\theta = -\frac{2\pi i}{3}(m+\frac{1}{2}) , \qquad m=0,2,3,5,\ldots \qquad (\lfloor m/3\rfloor + 1)^{th} \text{ order pole} \nonumber \\
            &\theta = \frac{2\pi i}{3}(m+2) , \qquad m=0,2,3,5,\ldots \qquad (\lfloor m/3\rfloor + 1)^{th} \text{ order pole}
        \end{align}
        where $\lfloor ~ \rfloor$ is the floor function and in third and fourth lines we exclude the values of $m$ when $ m\mod{} 3 = 1 $. We exclude these values of $m$ because, for $ m\mod{} 3 = 1 $, the poles obtained from $\theta = -\frac{2 \pi i}{3}(m+\frac{1}{2})$ coincide with those of $\theta = -(2m-1)\pi i$ and the poles obtained from $\theta = \frac{2 \pi i}{3}(m+2)$ coincide with those of $\theta = 2m\pi i$, hence increasing the order of these poles from $m$ to $2m$.

    \item {\it Zeroes}

        Zeroes of the dressing factor $\Phi_{DM}$ are given by the poles of gamma functions in the denominator and zeroes of the prefactor. Hence we have,
        \begin{align}
            &\theta = (2m+1)\pi i , \qquad m=1,2,3,\ldots \qquad 2m^{th} \text{ order zero} \nonumber \\
            &\theta = -2m\pi i , \qquad m=1,2,3,\ldots \qquad 2m^{th} \text{ order zero} \nonumber \\ 
            &\theta = -\frac{2\pi i}{3}(m+2) , \qquad m=0,2,3,5,\ldots \qquad (\lfloor m/3\rfloor + 1)^{th} \text{ order zero} \nonumber \\
            &\theta = \frac{2\pi i}{3}(m+\frac{7}{2}) , \qquad m=0,2,3,5,\ldots \qquad (\lfloor m/3\rfloor + 1)^{th} \text{ order zero} \nonumber \\
            &\theta = (2m - \frac{1}{2})\pi i , \qquad m\in \mathds{Z} \qquad \text{double zeroes}
        \end{align}
        where $\lfloor ~  \rfloor$ is the floor function and again in third and fourth lines we exclude the values of $m$ when $ m \mod 3 = 1 $. We do this again not because there are no zeroes at these values of $m$ but because the zeroes obtained from $\theta = -\frac{2 \pi i}{3}(m+2)$ coincide with those of $\theta = -2m\pi i$ and the zeroes obtained from $\theta = \frac{2 \pi i}{3}(m+\frac{7}{2})$ coincide with those of $\theta = (2m+1)\pi i$, hence increasing the order of these zeroes from $m$ to $2m$. The last set of zeroes are the zeroes of the prefactor $\sinh^2 \Big( \frac{i\pi}{4}+\frac{\theta}{2}\Big)$.
\end{itemize}
The poles and zeroes associated to the Fendley dressing phase $\Phi_F(\theta)$ are same as that of $\Phi_{DM}(\theta)$, except the zeroes coming from prefactor to the gamma functions. $\Phi_F(\theta)$ instead has double zeroes at $\theta = -(2m - \frac{1}{2})\pi i$, $m \in \mathds{Z}$. This means that $\Phi_F(\theta)$ has a zero at $\theta = \pi i/2$ which lies in the physical strip $\text{Im}(\theta) \in [0,\pi]$, which is not the case for $\Phi_{DM}(\theta)$.

With this assignment of poles and zeroes, and in particular with the absence of singularities in the physical strip, the De Martino - Moriconi dressing factor is essentially unique. The CDD ambiguity which is characteristic of integrable dressing factors is fixed in this case, since a CDD factor - see for instance \cite{Lonti} for the general expression in relativistic theories - would add a singularity in the strip Im$(\theta) \in [0,\pi]$. There is also a sign ambiguity which neither unitarity nor crossing can fix, however this is resolved by requiring $\Phi_{DM}(0)=+1$. Later on we shall connect with the $AdS_2$ dressing factor, and under the same assumptions of absence of singularities in the physical strip we can likewise expect the limiting $AdS_2$ dressing factor to therefore also be essentially unique. This uniqueness is to be contrasted with the fact that Fendley originally, and then De Martino and Moriconi in their improved analysis, found two different minimal $S$-matrices for the supersymmetric deformation of the tricritical Ising model. They are distinguished by the value of a parameter $p$, which can be either $\frac{1}{2}$ or $-\frac{3}{2}$. This difference however results in two different crossing equations for the associated dressing factors. A similar splitting is also reflected in the $AdS_2$ limit. We have focused on the $p=\frac{1}{2}$ case of eq. (\ref{eccola}) in this paper for definiteness.

The braiding unitarity condition also tends to $AdS_2$. In fact, as shown in \cite{AA2}, there is a combined braiding-unitarity condition that links the so-called solutions 3 and 5 of massless $AdS_2$. If one calls $\Omega_3$ and $\Omega_5$ the respective $AdS_2$ dressing factors, one finds
\begin{eqnarray}
\Omega_5(-\theta)\Omega_3(\theta) = \frac{1}{1+e^{-\theta}} = \frac{e^{ \frac{\theta}{2}}}{2\cosh \frac{\theta}{2}}.
\end{eqnarray}
This is precisely what descends from the condition 
\begin{eqnarray}
\Phi(-\theta)\Phi(\theta) = \frac{1}{1+\frac{ \sinh^2\frac{\theta}{2}}{\cosh^2 \theta}},
\end{eqnarray}
common to both $\Phi_F$ and $\Phi_{DM}$, in the limit $\theta \to \infty$. In fact, we have to consider that for $\theta \to \infty$ (and, consequently, $-\theta \to - \infty$) one should expect that
\begin{eqnarray}\label{Phi_expected_limits}
\Phi_A(\theta) \to \Omega_3(\theta), \qquad \Phi_A(-\theta) \to \Omega_5(-\theta), \qquad A=F,DM.  
\end{eqnarray}
Finally, physical unitarity works as well in the $AdS_2$ limit for both $\Phi_F$ and $\Phi_{DM}$, simply by considering (B.23) in \cite{AA2}.
 
These limiting considerations apply to the equations satisfied by the dressing factors. It is an interesting question whether either $\Phi_F$ or the new dressing phase of \cite{DM} really do tend as functions to the proposed $AdS_2$ dressing phase $\Omega(\theta)$ - formula (3.8) of \cite{AA1} and appendix B of \cite{AA2}. We will answer this question in the affirmative in a later section.
 
\section{Minimal Two-Particle Form Factors}

Let us now use the knowledge of the $S$-matrix data to construct solutions to the form-factor axioms, focusing on the two-particle case.

If we recall that the $S$-matrix is related to the $R$-matrix as
\begin{eqnarray}
S = \Pi \circ R
\end{eqnarray}
with $\Pi$ the graded permutation, we can easily obtain the $S$-matrix entries. Let us start with the processes involving in-states (indices $a,b$ in $S_{ab}^{cd}$) and out-states (indices $c,d$ in $S_{ab}^{cd}$) such that $a+b=c+d=0$ (mod 2). This means that the in- and out- states have total fermion number zero, and so can only be $bb$ or $ff$:
\begin{eqnarray}
S_{bb}^{bb}=\Phi_F(\theta)= S_{ff}^{ff}, \quad S_{bb}^{ff} =  -S_{ff}^{bb}=\Phi_F(\theta)\mbox{sech} \theta \, \sinh \frac{\theta}{2}.
\end{eqnarray}
We also recall that any $S$-matrix will preserve the total fermion number mod 2. 

A useful strategy for how to proceed, if we are interested for instance in the two-particle minimal form factors, can be found in the paper \cite{Mussardo:1998kq}. The first thing to do is to (flavour-)diagonalise the $S$-matrix. One can check that the (un-normalised) bosonic (flavour-)eigenstates are given by
\begin{eqnarray}
|b_\pm\rangle = |b\rangle \otimes |b\rangle \pm i  |f\rangle \otimes |f\rangle,
\end{eqnarray}   
with eigenvalues 
\begin{eqnarray}
\Phi_F(\theta)\lambda_\pm (\theta)= \Phi_F(\theta)\Big(1 \mp i \, \mbox{sech} \theta \, \sinh \frac{\theta}{2}\Big),
\end{eqnarray}
respectively.
We remind that the tensor product is graded, hence
\begin{eqnarray}
A \otimes B  |i\rangle \otimes |j\rangle= (-)^{|B| |i|} A|i\rangle \otimes B|j\rangle,
\end{eqnarray}
and that $|b\equiv 1|=0$, $|f\equiv 2|=1$, $|E_{ij}|=|i|+|j|$ mod 2. Likewise, using the fact that
\begin{eqnarray}
S_{bf}^{fb} = -S_{fb}^{bf}=-\Phi_F(\theta)\tanh \theta, \qquad S_{bf}^{bf} = S_{fb}^{fb}=\Phi_F(\theta)\mbox{sech}\theta \, \cosh \frac{\theta}{2},
\end{eqnarray}
we find that the (un-normalised) fermionic (flavour-)eigenstates are given by
\begin{eqnarray}
|f_\pm\rangle = |b\rangle \otimes |f\rangle \pm i  |f\rangle \otimes |b\rangle,
\end{eqnarray}   
with eigenvalues 
\begin{eqnarray}
\Phi_F(\theta)\mu_\pm (\theta)= \Phi_F(\theta)\Big(\mbox{sech} \theta \, \cosh \frac{\theta}{2} \pm i \tanh \theta\Big),
\end{eqnarray}
respectively.
For later convenience, let us divide the contributions to the $S$ matrix into a redefined dressing phase for the fermionic block:
\begin{eqnarray}
\Phi_n (\theta) \equiv \mu_+(\theta) \Phi_F(\theta), \qquad \sigma_{ab}^{cd} \equiv \frac{S_{ab}^{cd}}{\Phi_n(\theta)},  \label{splitto}
\end{eqnarray}
and try to solve the permutation axiom separately for the new dressing part $\Phi_n$ and the non-dressing part $\sigma_{ab}^{cd}$. Let us begin with the latter and consider the case of $2$-particle form factors. The $S$-matrix is of course diagonal on the eigenstates. Therefore, we define the form factors 
\begin{eqnarray}
B^{tot}_\pm \equiv \langle0|{\cal{O}}_b|b_\pm\rangle, \qquad F^{tot}_\pm \equiv \langle0|{\cal{O}}_f|f_\pm\rangle, \label{ket}
\end{eqnarray}
with ${\cal{O}}_b$ being a bosonic operator and ${\cal{O}}_f$ being a fermionic operator, both operators though of spin $0$ for simplicity (for the fermion, this is the case for instance of the fictitious operator introduced in \cite{Mussardo:1998kq}, around formula (6.13) and the preceding discussion). We first focus on the fermionic form factors, then the permutation axiom just becomes (for the non-dressing part)
\begin{eqnarray}
&&F_+(\theta) = F_+(-\theta), \qquad F_-(\theta) = \frac{\mu_-(\theta)}{\mu_+(\theta)}F_-(-\theta),\label{permufo}
\end{eqnarray}
while the periodicity axiom just reads
\begin{eqnarray}
F_+(\theta+2\pi i) = -i F_+(-\theta), \qquad F_-(\theta+2\pi i) = i F_-(-\theta).\label{periofo}
\end{eqnarray} 
The purely difference form with $\theta = \theta_1-\theta_2$ ($\theta_1$ being the rapidity of the first particle in the ket in (\ref{ket}), and $\theta_2$ the rapidity of the second, since $|b_\pm\rangle$ and $|f_\pm\rangle$ are two-particle states) is due to Lorentz invariance (with the caveat of the spin being zero for both the bosonic and the fermionic operator, see comment above).
This form of periodicity descends from the fact that cycling the states we get
\begin{eqnarray}
|f_\pm\rangle = |b\rangle \otimes |f\rangle \pm i  |f\rangle \otimes |b\rangle |\to (cycling) \to |f\rangle \otimes |b\rangle \mp i  |b\rangle \otimes |f\rangle = \mp i |f_\pm\rangle.
\end{eqnarray} 
There is a {\it caveat}: this is of course assuming that the statistical factor is just $\pm 1$, while we know that it might be more complicated (semi-locality index). This means that we are at this moment searching for a solution where the statistical factors are the standard ones.

It is important to note that the splitting (\ref{splitto}) is crucial if we want to solve separately the dressing and non-dressing parts of the form factor axioms. In fact, thanks to this choice, we have that (\ref{permufo}) is not manifestly inconsistent: iterating it produces
\begin{eqnarray}
&&F_+(-\theta) = F_+(\theta) = F_+(-\theta),\nonumber\\
&&F_-(-\theta) = \frac{\mu_-(-\theta)}{\mu_+(-\theta)}F_-(\theta) = \frac{\mu_-(-\theta)}{\mu_+(-\theta)}\frac{\mu_-(\theta)}{\mu_+(\theta)}F_-(-\theta),
\end{eqnarray}
which is consistent because, by inspection, we have
\begin{eqnarray}
\mu_\pm(\theta) = \mu_\mp(-\theta).
\end{eqnarray}
This means that we can hope to solve the non-dressing part (formulated in terms of the eigenstates) and  the dressing part separately. The form factor pertaining to the dressing phase will solve
\begin{eqnarray}
F^{dr}(\theta) = \Phi_n(\theta)F^{dr}(-\theta), \qquad F^{dr}(\theta+2\pi i) = F^{dr}(-\theta),\label{notar}
\end{eqnarray} 
such that 
\begin{eqnarray}
F^{tot}_\pm(\theta) = F^{dr}(\theta)F_\pm(\theta)
\end{eqnarray}
is a solution to the axioms of a two-particle form factor. If we require this object to have the least possible number of singularities, we can claim that we have found the {\it minimal} two-particle form factor. Notice that (\ref{notar}) is not immediately inconsistent either, since it implies
\begin{eqnarray}
\Phi_n(-\theta) = \frac{F^{dr}(-\theta)}{F^{dr}(\theta)} = \frac{1}{\Phi_n(\theta)},
\end{eqnarray} 
which is true. In fact, from the properties of $\mu_\pm$ and the braiding unitarity property of $\Phi_F$, we have
\begin{eqnarray}
\Phi_n(-\theta) = \mu_+(-\theta) \Phi_F(-\theta) = \frac{\mu_-(\theta)}{\Phi_F(\theta) \Big[1+\frac{\sinh^2 \frac{\theta}{2}}{\cosh^2 \theta}\Big]} = \frac{1}{\Phi_F(\theta)\mu_+(\theta)} = \frac{1}{\Phi_n(\theta)},\label{same}
\end{eqnarray}
where the second to last step just involves identities of hyperbolic functions.

The equation for the $B_\pm$ form factors are in a way simpler. In fact periodicity implies that cycling a bosonic operator gives
\begin{eqnarray}
|b_\pm\rangle = |b\rangle \otimes |b\rangle \pm i  |f\rangle \otimes |f\rangle |\to (cycling) \to |b\rangle \otimes |b\rangle \pm i  |f\rangle \otimes |f\rangle = |b_\pm\rangle.
\end{eqnarray} 
In order to have once again a consistent splitting between dressing and non-dressing parts (the latter being again given in terms of the eigenbasis), we make a different choice in the bosonic {\it vs.} the fermionic block:
\begin{eqnarray}
&&\qquad \qquad \quad \,\,\,\,\, |b_+\rangle
\qquad \qquad \qquad \quad \,\,\, |b_-\rangle
\qquad \qquad \qquad \quad \,\,\, |f_+\rangle
\qquad \qquad \qquad \quad \,\,\,\, |f_-\rangle\qquad \qquad\nonumber\\
&&S = \begin{pmatrix}[\Phi_F(\theta)\lambda_+(\theta)] \times 1&0&0&0\\0&[\Phi_F(\theta)\lambda_+(\theta)] \times \frac{\lambda_-(\theta)}{\lambda_+(\theta)}&0&0\\0&0&[\Phi_F(\theta)\mu_+(\theta)]\times 1&0\\0&0&0&[\Phi_F(\theta)\mu_+(\theta)] \times \frac{\mu_-(\theta)}{\mu_+(\theta)}\end{pmatrix}.\nonumber
\end{eqnarray}
This splitting is consistent since it produces
\begin{eqnarray}
&&B_+(\theta) = B_+(-\theta), \qquad B_-(\theta) = \frac{\lambda_-(\theta)}{\lambda_+(\theta)}B_-(-\theta),\label{permuf}
\end{eqnarray}
from permutation, with $\lambda_\pm(\theta) = \lambda_\mp(-\theta)$, while the periodicity axiom reads
\begin{eqnarray}
B_+(\theta+2\pi i) = B_+(-\theta), \qquad B_-(\theta+2\pi i) = B_-(-\theta).\label{periof}
\end{eqnarray} 
The form factor pertaining to the dressing phase in the bosonic block will solve
\begin{eqnarray}
B^{dr}(\theta) = \Phi_m(\theta)B^{dr}(-\theta), \qquad B^{dr}(\theta+2\pi i) = B^{dr}(-\theta), \qquad \Phi_m(\theta) \equiv \Phi_F(\theta)\lambda_+(\theta),\label{fre}
\end{eqnarray} 
such that 
\begin{eqnarray}
B^{tot}_\pm(\theta) = B^{dr}(\theta)B_\pm(\theta)
\end{eqnarray}
is a solution to the axioms of a two-particle form factor (which can again be chosen to be minimal). The first relation in (\ref{fre}) is also not manifestly inconsistent: one can repeat the same argument as in (\ref{same}), with $\lambda$ replacing $\mu$ everywhere, and it will still work. Since these considerations only rely on the braiding unitarity equation, they in fact hold both for $\Phi_F$ and for $\Phi_{DM}$.

\subsection{Solution for the dressing phase \label{tech}}

We shall solve the equations for the dressing factor by resorting to the Karowski-Weisz theorem \cite{a1} - see also footnote 1 in \cite{Ale}. This can be applied irrespectively to the Fendley or the De Martino - Moriconi dressing phase, as it only relies on the property (\ref{same}):
\begin{eqnarray}
\Phi_n(-\theta) = \frac{1}{\Phi_n(\theta)}.
\end{eqnarray} 

The idea is to recast the dressing factor as
\begin{eqnarray}
\Phi_n(\theta) = \exp \int_0^\infty dt \, f(t) \sinh \frac{\theta t}{i \pi}.\label{reca}
\end{eqnarray}
We can try to solve (ignoring issues of branches for the moment)
\begin{eqnarray}
\log[\Phi_n(\theta)]= \int_0^\infty dt \, f(t) \sinh \frac{\theta t}{i \pi} = -\frac{1}{2}\int_0^\infty dt \, f(t) \, e^{\frac{i\theta t}{\pi}}+\frac{1}{2}\int_0^\infty dt \, f(t) \, e^{-\frac{i\theta t}{\pi}}.\label{rihs}
\end{eqnarray}
We now make a working assumption, which will need to be verified at the very end by looking at the solution which we will find, that
\begin{eqnarray}
f(t) = -f(-t).\label{simo}
\end{eqnarray}
This allows us to change variable in the second integral and to rewrite the rhs of (\ref{rihs}) as a single integral:
\begin{eqnarray}
\log[\Phi_n(\theta)] =-\frac{1}{2}\int_{-\infty}^\infty dt f(t) \, e^{\frac{i\theta t}{\pi}}.
\end{eqnarray}
This is simply a statement of Fourier transforms: 
\begin{eqnarray}
\log [\Phi_n(\theta)]=-\sqrt{\frac{\pi}{2}}\tilde{f}\Big(\frac{\theta}{\pi}\Big), \qquad \tilde{f}(k) = \frac{1}{\sqrt{2\pi}}\int_{-\infty}^\infty dt f(t) \, e^{ik t}.
\end{eqnarray}
We can simply invert the transform and extract $f(t)$, by writing
\begin{eqnarray}
f(t) = \frac{1}{\sqrt{2\pi}}\int_{-\infty}^\infty dk \, \tilde{f}(k) \, e^{-ik t} = -\frac{1}{\pi} \int_{-\infty}^\infty d\theta \, \log[ \Phi_n(\pi \theta)] \, e^{- i \theta t}. \label{effe}
\end{eqnarray}
We now need to check the consistency of our working assumption, which involved $f(t)$ being odd. Indeed, this relies on (\ref{same}), since
\begin{equation}
\begin{aligned}
f(-t) & =  -\frac{1}{\pi} \int_{-\infty}^\infty d\theta \, \log [\Phi_n(\pi \theta)] \, e^{i \theta t} = -\frac{1}{\pi} \int_{-\infty}^\infty d\theta \, \log [\Phi_n(- \pi \theta)] \, e^{-i \theta t} 
\\
& = \frac{1}{\pi} \int_{-\infty}^\infty d\theta \, \log [\Phi_n(\pi \theta)] \, e^{-i \theta t}= - f(t)
\end{aligned}
\end{equation}
where we have changed variable at the second step, and used (\ref{same}) at the third step. 

This means that we can simply apply the Karowski-Weisz theorem, which says that one solution to (\ref{notar}) is given by
\begin{eqnarray}
F^{dr}(\theta) = \exp \frac{1}{2} \int_{-\infty}^\infty dt \, f(t) \frac{\sin^2{[\frac{t(i\pi - \theta)}{2\pi}]}}{\sinh t},
\end{eqnarray}
with $f(t)$ given by (\ref{effe}), and where we have extended the integral using the even parity of $\frac{f(t)}{\sinh t}$. Notice that we could also multiply $F^{dr}(\theta)$ by an arbitrary constant and it would still be a solution. A completely analogous argument works for the solution to (\ref{fre}), by simply replacing $\Phi_n$ by $\Phi_m$ everywhere it appears, and $F$ with $B$.

It is important to comment on the validity of the procedure which we have just outlined. The manipulations of the integral which we have performed are formal, and require an analysis of the conditions under which they can be carried out. This is particularly important, since we will soon work out, as a toy example, the computation for a simple choice of $\Phi_{n}(\theta)$. This leads to an explicit expression for $f(t)$ provided the integrals in the intermediate steps are understood in the distributional sense. The conditions which we require in the first instance are that the $\log[\Phi_n(\theta)]$ part of the integrand does not grow at $|\theta| \to \infty$ faster than a polynomial, {\it i.e.} are a tempered distribution. In this way, the Fourier transform producing $f$ will also be a tempered distribution. The conclusion of our procedure involves then integrating $f$ against a rapidly decreasing function of the form $\sim \frac{\sin^2 t}{\sinh t}$, which behaves like a test function and guarantees a finite result when integrated with a tempered distribution. In the situation where the integrand should evade this assumption, we might still need to extend our method but we would then need to adopt a suitable regularisation, as is the case in other more traditional approaches to form factors  \cite{BabuF,Babujian:2001xn}. In those instances, such a regularisation will have to be ascertained on a case by case basis via a detailed study of the integrand - in \cite{Babujian:2001xn} for instance the idea of analytically  continuing in the potential parameters appearing (such as what they define as $\nu$ for Sine-Gordon in their specific case) to achieve convergence of the integrals is discussed.

As anticipated, we shall now work out the computation in the following simple case:
\begin{equation}
A(\theta) = e^{\frac{\theta}{2}}A(-\theta) \qquad \qquad A(\theta + 2\pi i)=A(-\theta) \,\, .
\end{equation}
It is clear that the function $e^{\frac{\theta}{2}}$ goes into its reciprocal under a change of sign of $\theta$ and one could actually just guess the solution by simple inspection:
\begin{eqnarray}
A(\theta) = c \, e^{\frac{\theta}{4}+i\frac{\theta^2}{8\pi}},\label{inspe}
\end{eqnarray}
with $c$ any constant.
However notice that, although in a slightly unorthodox way, our method does work in this case very explicitly. In fact, we would then simply say 
\begin{eqnarray}
A(\theta) = \exp \frac{1}{2} \int_{-\infty}^\infty dt \, a(t) \frac{\sin^2{[\frac{t(i\pi - \theta)}{2\pi}]}}{\sinh t},
\end{eqnarray}
with
\begin{eqnarray}
a(t) = -\frac{1}{\pi} \int_{-\infty}^\infty d\theta \, \log{[e^{\frac{\pi \theta}{2}}]} \, e^{- i \theta t} = - \pi i \, \frac{d}{dt} \delta(t).
\end{eqnarray}
This implies 
\begin{eqnarray}
A(\theta) = \exp \Bigg[- i \frac{\pi}{2} \int_0^\infty dt \, \frac{d}{dt} \delta(t) \frac{\sin^2{[\frac{t(i\pi - \theta)}{2\pi}]}}{\sinh t}\Bigg] = \exp i \frac{\pi}{2} \Bigg[\frac{d}{dt} \frac{\sin^2{[\frac{t(i\pi - \theta)}{2\pi}]}}{\sinh t}\Bigg]_{t=0} = \exp \Big[- \frac{i}{8} + \frac{\theta}{4} + i \frac{\theta^2}{8\pi}\Big].\nonumber
\end{eqnarray}
Considering that a constant can be fixed arbitrarily, we get exactly the solution which we guessed by inspection in (\ref{inspe}). With this explicit example we have therefore derived a lesson, concerning the applicability of our technique. That is, the convergence of the integral involved might have to be understood in the \underline{distributional} sense. If one accepts that, as we have seen, this seems to  produce a finite result, which one can then check in those cases where an explicit solution is otherwise available.

\subsection{Eigenvalues part}

In fact, one can go further: the same method solves the equations for $B_-$ in (\ref{permuf}) and (\ref{periof}), given that 
\begin{eqnarray}
\frac{\lambda_-(\theta)}{\lambda_+(\theta)} \equiv \Lambda(\theta)
\end{eqnarray}
again satisfies the key relation which will ensure the consistency of the procedure:
\begin{eqnarray}
\Lambda(-\theta) = \frac{1}{\Lambda(\theta)}.
\end{eqnarray}
Hence the formal solution for $B_-$ is, by the very same argument outlined in the previous section,
\begin{eqnarray}
B_-(\theta) = \exp \frac{1}{2} \int_{-\infty}^\infty dt \, g_-(t) \frac{\sin^2{[\frac{t(i\pi - \theta)}{2\pi}]}}{\sinh t},
\end{eqnarray}  
with
\begin{eqnarray}
g_-(t) = -\frac{1}{\pi} \int_{-\infty}^\infty d\theta \, \log [\Lambda(\pi \theta)] \, e^{- i \theta t}.
\end{eqnarray}
Of course a consistent solution for $B_+$ is just a constant solution $B_+(\theta) = const.$, as one can see from (\ref{permuf}) and (\ref{periof}). The method introduced in the previous section would also produce this solution via a $\log 1 = 0$ as the function $f$.

In the case of $F_+$, from (\ref{permufo}) and (\ref{periofo}) we see that we can adopt the method above, with a slight refinement first. Let us take the equation
\begin{eqnarray} \label{F+axioms}
F_+(\theta + 2 \pi i) = -i \, F_+(-\theta) = - i \, F_+(\theta),
\end{eqnarray}
where we have combined the two conditions on $F_+$. Let us define
\begin{eqnarray}
\tilde{F}_+(\theta) \equiv u(\theta) \, F_+(\theta),\label{defi}
\end{eqnarray}
and look for an appropriate $u(\theta)$ leading, via \eqref{F+axioms}, to the conditions
\begin{eqnarray}
\tilde{F}_{+}(\theta)=\phi(\theta) \tilde{F}_{+}(-\theta)
\qquad \qquad
\tilde{F}_+(\theta + 2 \pi i) = \tilde{F}_+ (-\theta).  \label{karo1}
\end{eqnarray}
As shown in the previous section, these can be solved via the Karowski-Weisz theorem provided that $\phi(-\theta)=\frac{1}{\phi(\theta)}$. Given the solution one can then simply extract $F_{+}(\theta)$ from \eqref{defi}. It turns out that one can obtain \eqref{karo1} provided that $u(\theta)$ satisfies the relation
\begin{equation}\label{u-requirement}
u(\theta+2\pi i)=iu(-\theta)
\end{equation}
and $\phi(\theta)$ is defined as
\begin{equation}
\phi(\theta)\equiv \frac{u(\theta)}{u(-\theta)}\,\, ,
\end{equation}
which clearly respects the condition $\phi(-\theta)=\frac{1}{\phi(\theta)}$.
The axioms for $F_{+}(\theta)$ can then be solved by finding $u(\theta)$ satisfying \eqref{u-requirement}. We start by noting that
\begin{equation}
\frac{u(\theta+2\pi i)}{u(-\theta)}=i \qquad \xrightarrow[\theta \rightarrow \theta-\pi i]{} \qquad
\frac{u(\theta + \pi i)}{u(-\theta+i\pi)} = i \qquad \xrightarrow[\theta \rightarrow -\theta]{} \qquad \frac{u(-\theta + \pi i)}{u(\theta+i\pi)} = i \, \,  \,\, ,
\end{equation}
which shows inconsistency of \eqref{u-requirement} for meromorphic functions and hints toward the fact that $u(\theta)$ should exhibit branch cuts. We then propose the following piecewise-defined function
\begin{equation}
u(\theta) = 
\begin{cases}
B_1\sqrt{-\sinh{(\theta - i \epsilon)}} \qquad \text{for} \qquad Re[\theta] >0,
\\
B_2 \sqrt{\sinh{(\theta+i \epsilon)}} \qquad\,\,\,\,\,\, \text{for} \qquad Re[\theta] < 0,
\end{cases}
\end{equation}
with $B_1$ and $B_2$ arbitrary constants, which solves the desired condition when $\theta$ is real and indeed respects the above expectation. Another possible solution might be obtained by replacing $\sinh{\theta}$ with $\tanh{\theta}$. The lack of analyticity of $u(\theta)$ could be ascribed to a number of factors. One possibility would be that the dressing-factor part, which we have expressed as an integral, might in principle have a similar non-analyticity which may cancel it. We have not extensively analysed that aspect, which is left for future work. Alternatively, our assignment of statistical factors in the periodicity axiom might differ upon the choice of the operators, and it might actually alter the periodicity condition. Furthermore, one meromorphic solution to all the axioms is a vanishing two-particle form factor ({\it i.e.} $B_1=B_2=0$). It is still logically possible that the theory might have a super-selection principle which for instance makes all the even-point functions vanish. Finally, it could be that the types of $S$-matrices which can be related to $AdS_2$ present a non-analytic solution, which would be yet another one of the non-standard features of these theories.

Finally, we can do the more complicated case of $F_-$ with a slightly more elaborated trick, which reduces the problem, once more, to the same structure. Consider (\ref{permufo}) and (\ref{periofo}), and define
\begin{eqnarray}
G_-(\theta) \equiv \mu_+(\theta) F_-(\theta). \label{defii}
\end{eqnarray}
Using the fact that $\mu_\pm(\theta) = \mu_\mp(-\theta)$ we can recast the second equation in (\ref{permufo}) as
\begin{eqnarray}
\mu_+(\theta)F_-(\theta) =\mu_+(-\theta)F_-(-\theta), \qquad \mbox{namely} \qquad G_-(\theta) = G_-(-\theta).
\end{eqnarray}
The second equation in (\ref{periofo}) then implies
\begin{eqnarray}
G(\theta+2\pi i) = \mu_+(\theta+2 \pi i)F_-(\theta+2\pi i) = - \mu_+(-\theta) \, i \, F_-(-\theta) = -i \, G_-(-\theta).
\end{eqnarray}
In this derivation we have also used the fact that $\mu_+(\theta+2 \pi i) = - \mu_+(-\theta)$. This means that $G_-$ satisfies the exact same system of equations as $F_+$, therefore we can take the solution {proposed above upon the introduction of $\tilde{F}_{+}(\theta)$ and successively reconstruct $F_-$ by using (\ref{defii}).

\section{Numerics Of The De Martino - Moriconi Dressing Factor}
In this section we use a numerical approach to study the behaviour of the integral representation \eqref{inteo} of the improved dressing phase proposed by De Martino and Moriconi, providing evidence for \eqref{Phi_expected_limits}. To study the limits $\theta \rightarrow \pm \infty$ of the DM and $AdS_2$ dressing phases, we begin by rewriting them as
\begin{equation}
\begin{aligned}
\Phi_{DM}(\theta) = \exp \Bigg[-\int_0^\infty dt \, f(t,\theta)\Bigg] \qquad \qquad \qquad \, &\text{with} \qquad f(t,\theta)\equiv h(t) \, \frac{\sin t \theta \, \sin{[t(i\pi-\theta)]}}{t \cosh \pi t},
\\
\Omega_{3}(\theta)= \frac{e^{-i\tfrac{\pi}{8}}}{\sqrt{2}} \exp \Biggl[ \frac{\theta}{4}+\frac{1}{2} \int_{0}^{\infty} dx \, l(x,\theta)
 \Biggr] \qquad &\text{with} \qquad \, l(x,\theta)\equiv \frac{\cosh{[x(1-\tfrac{2\theta}{i\pi})]}-\cosh{x}}{x \cosh{x}\sinh{2x}},
\end{aligned}
\end{equation}
\begin{eqnarray}\label{omega5_function_of_omega3}
\Omega_{5}(\theta)= \frac{1}{\Omega_{3}(-\theta)}\frac{1}{1+e^{\theta}} \, \, ,
\end{eqnarray}
with $h(t)$ defined in \eqref{inteo} and $\Omega_{3}(\theta),\Omega_{5}(\theta)$ respectively given in equations (B.28) and (B.4) of \cite{AA2}. In order to compare the two dressing factors for large positive and negative values of $\theta$, one needs to numerically approximate the integrals appearing in the exponentials and to this aim it is useful to perform a preliminary study of the two integrands $f(t,\theta),l(x,\theta)$, so as to identify an appropriate restriction for the integration domains. For increasing positive and negative values of $\theta$ one finds the following behaviours
\begin{figure}[H]
\begin{minipage}{0.48\textwidth}
\centering
\includegraphics[width=1\linewidth]{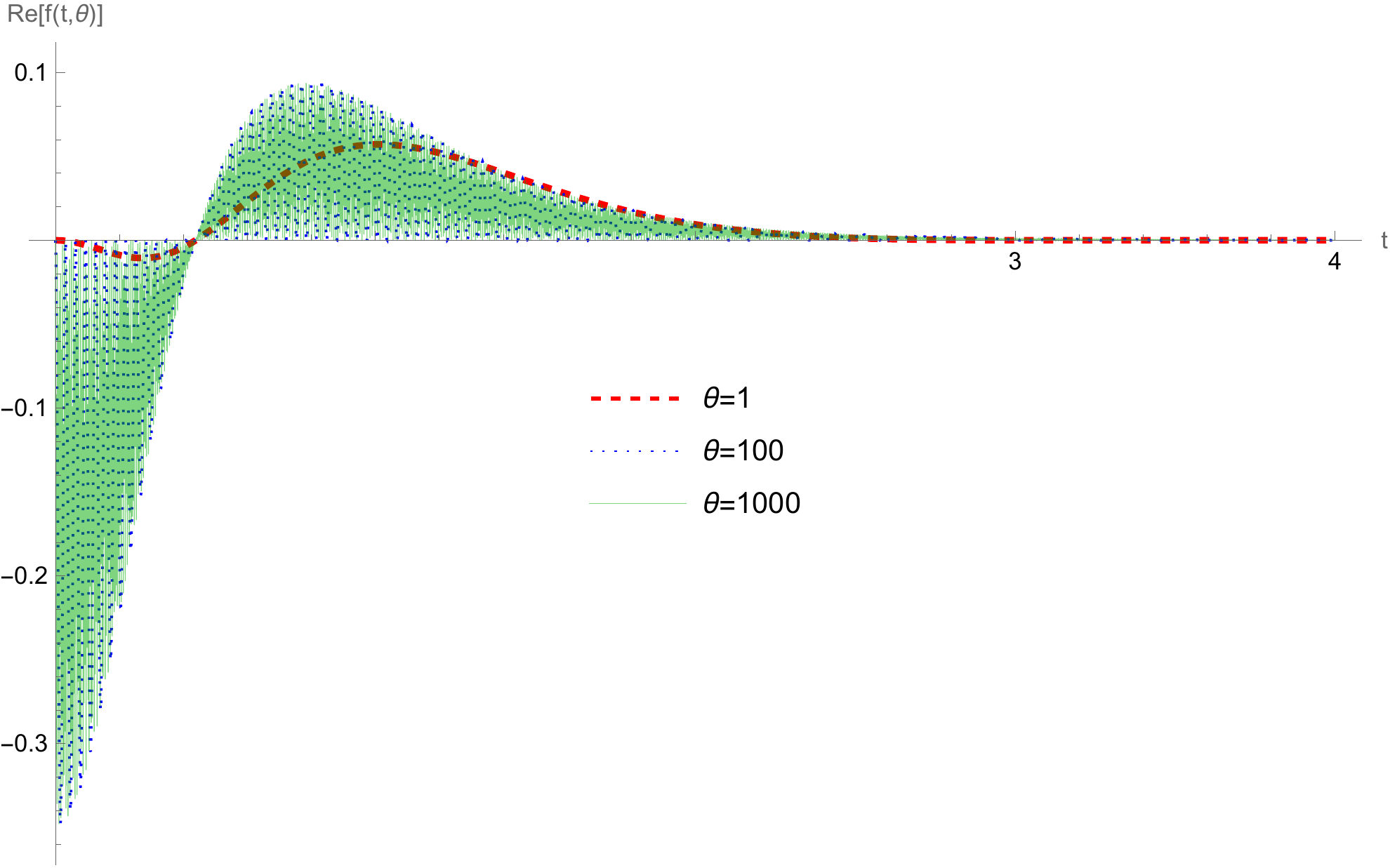}
\caption{Real part of $f(t,\theta)$ for $\theta\rightarrow +\infty$}\label{Fig:Re_f_theta+}
\end{minipage}\hfill
\begin{minipage}{0.48\textwidth}
\centering
\includegraphics[width=1\linewidth]{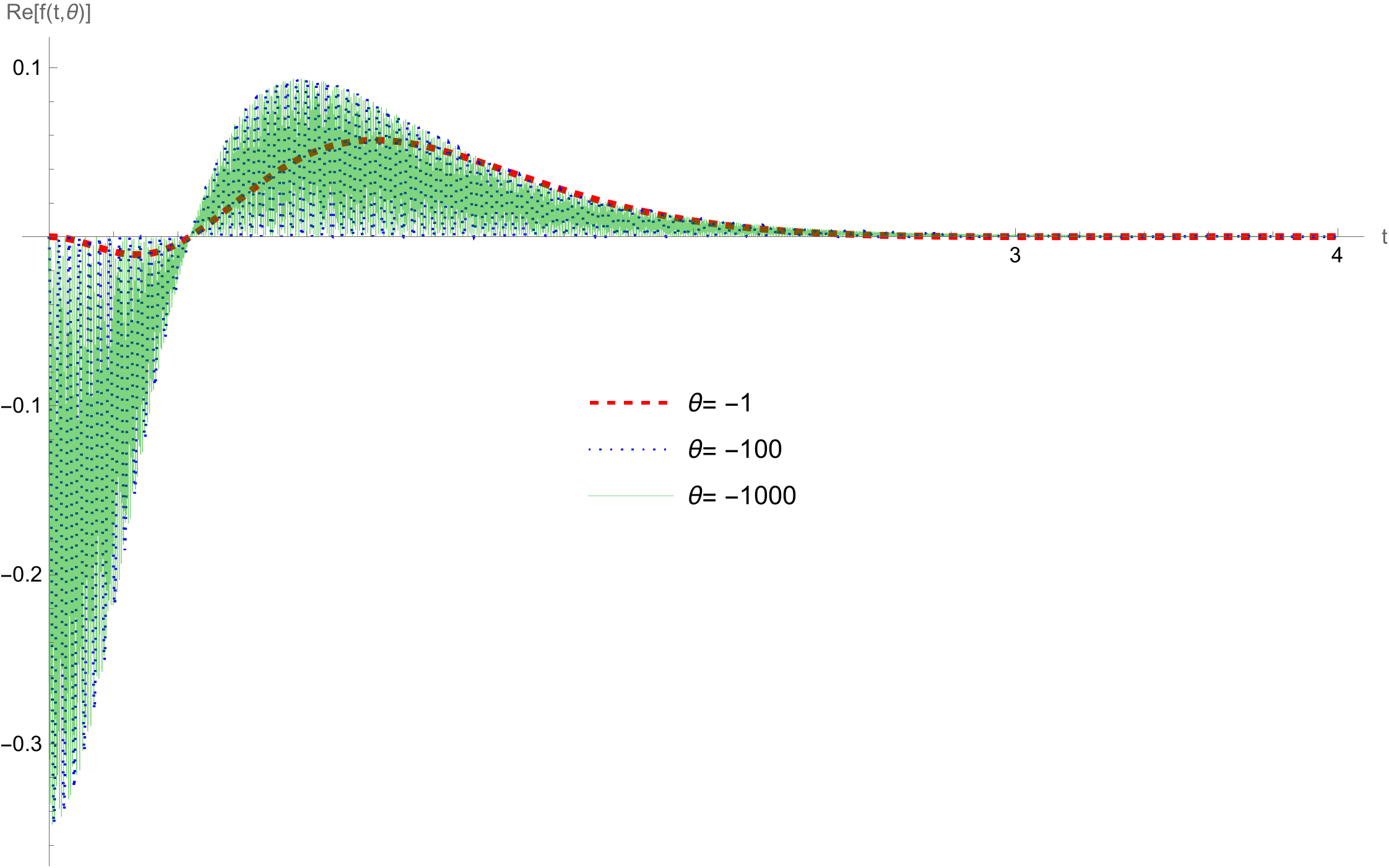}
\caption{Real part of $f(t,\theta)$ for $\theta\rightarrow -\infty$}\label{Fig:Re_f_theta-}
\end{minipage}
\end{figure}
\begin{figure}[H]
\begin{minipage}{0.48\textwidth}
\centering
\includegraphics[width=1\linewidth]{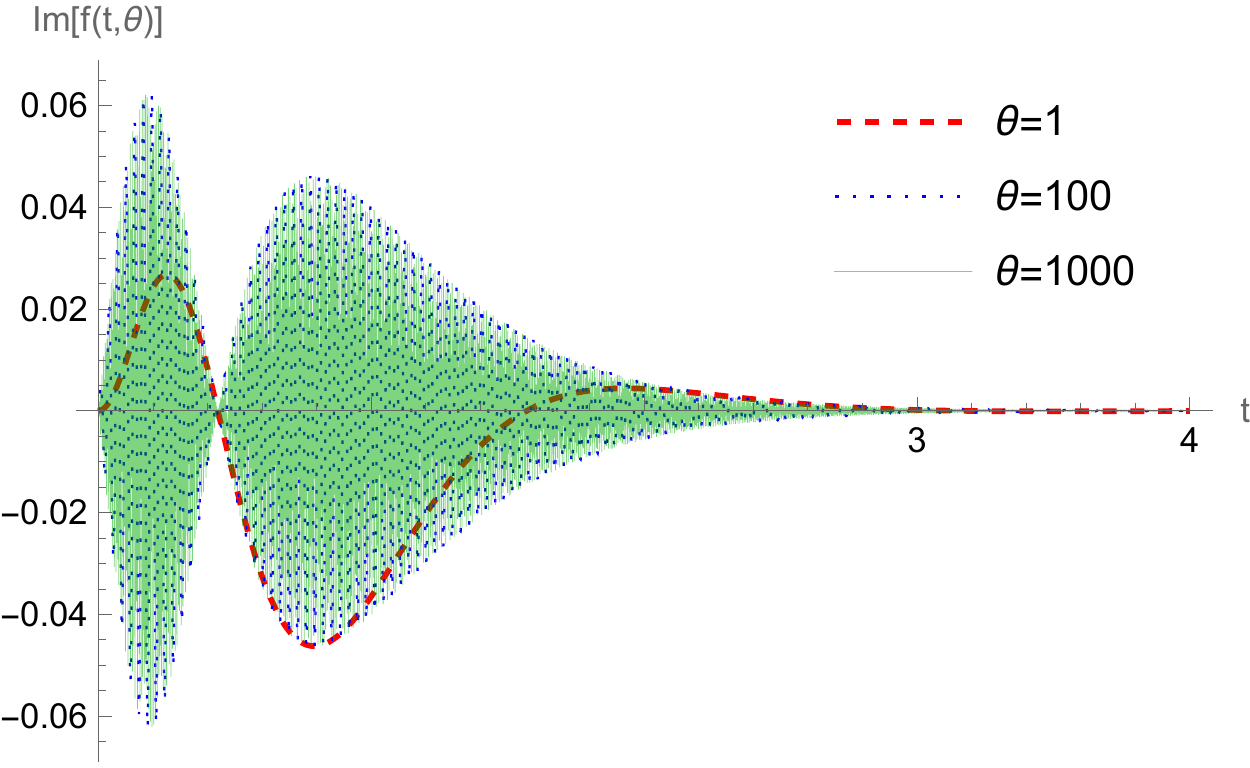}
\caption{Imaginary part of $f(t,\theta)$ for $\theta\rightarrow +\infty$}\label{Fig:Im_f_theta+}
\end{minipage}\hfill
\begin{minipage}{0.48\textwidth}
\centering
\includegraphics[width=1\linewidth]{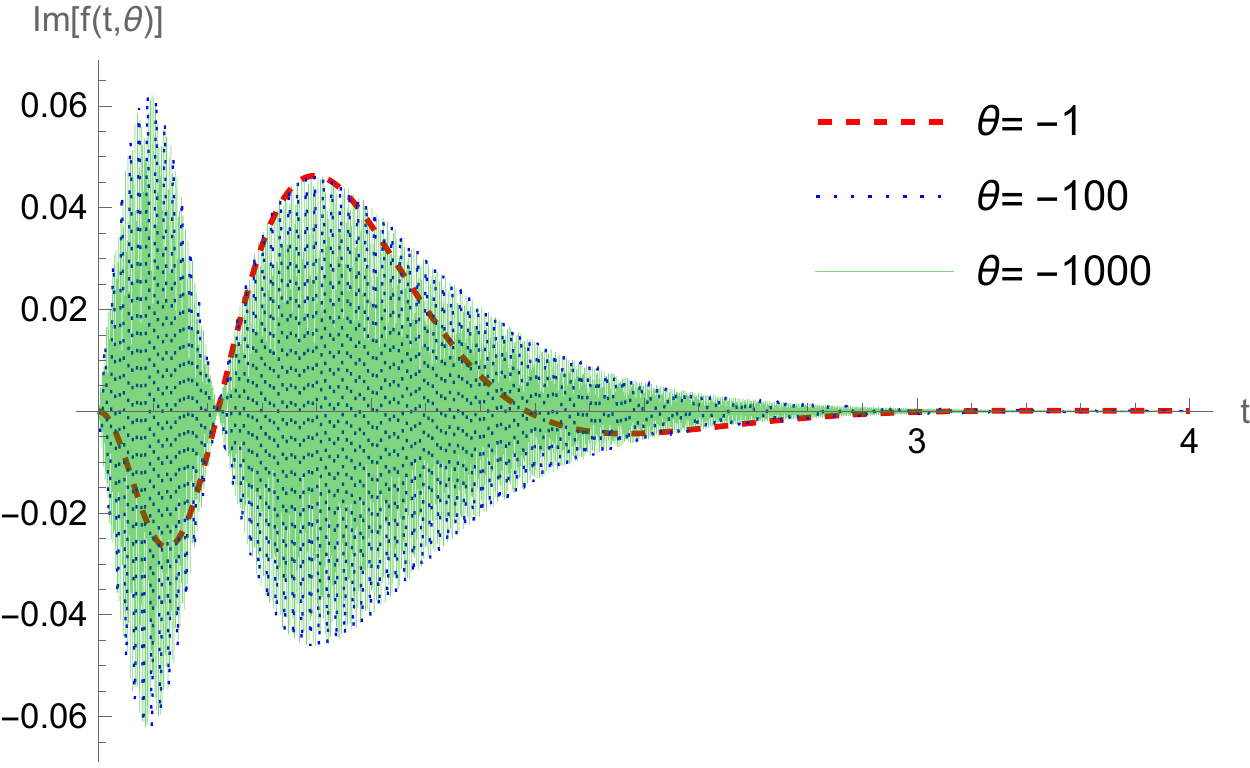}
\caption{Imaginary part of $f(t,\theta)$ for $\theta\rightarrow -\infty$}\label{Fig:Im_f_theta-}
\end{minipage}
\end{figure}
\begin{figure}[H]
\begin{minipage}{0.48\textwidth}
\centering
\includegraphics[width=1\linewidth]{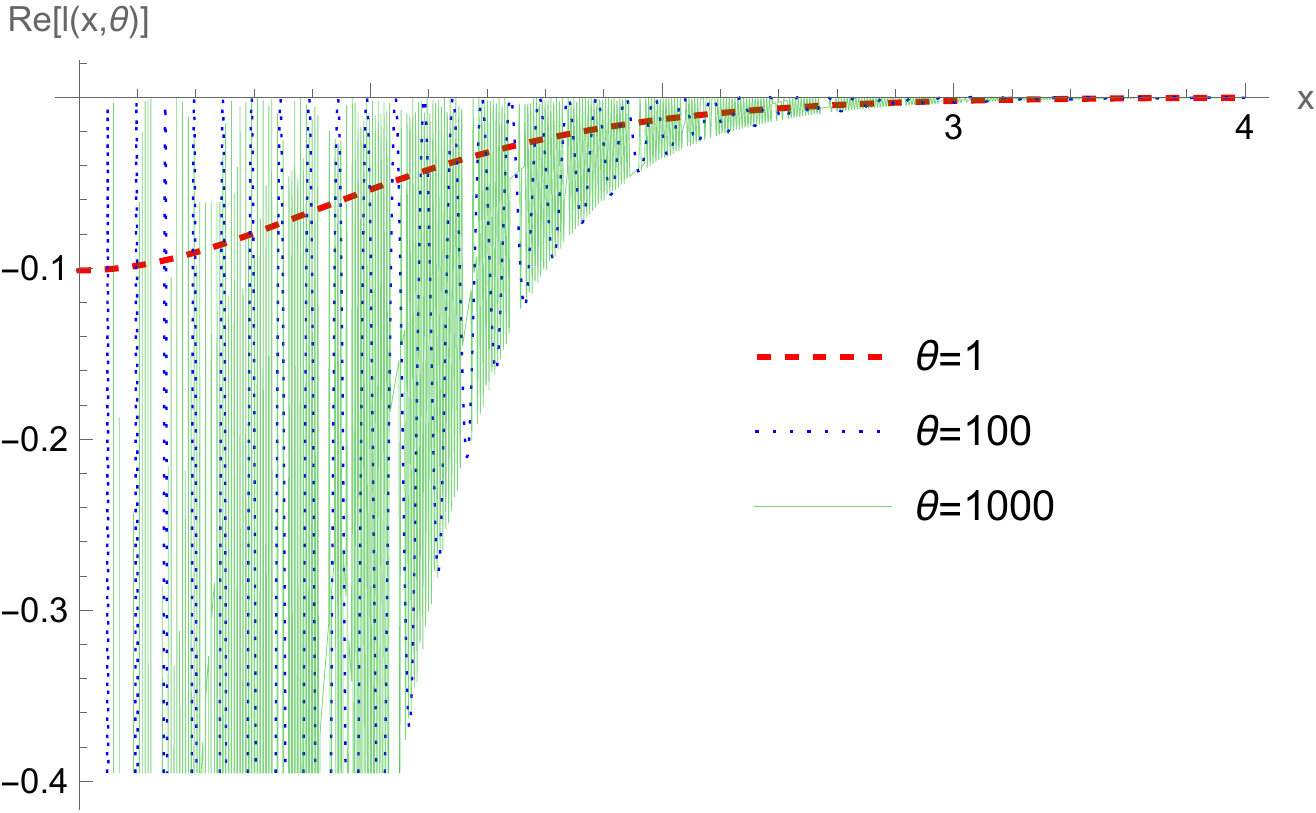}
\caption{Real part of $l(x,\theta)$ for $\theta\rightarrow +\infty$}\label{Fig:Re_l_theta+}
\end{minipage}\hfill
\begin{minipage}{0.48\textwidth}
\centering
\includegraphics[width=1\linewidth]{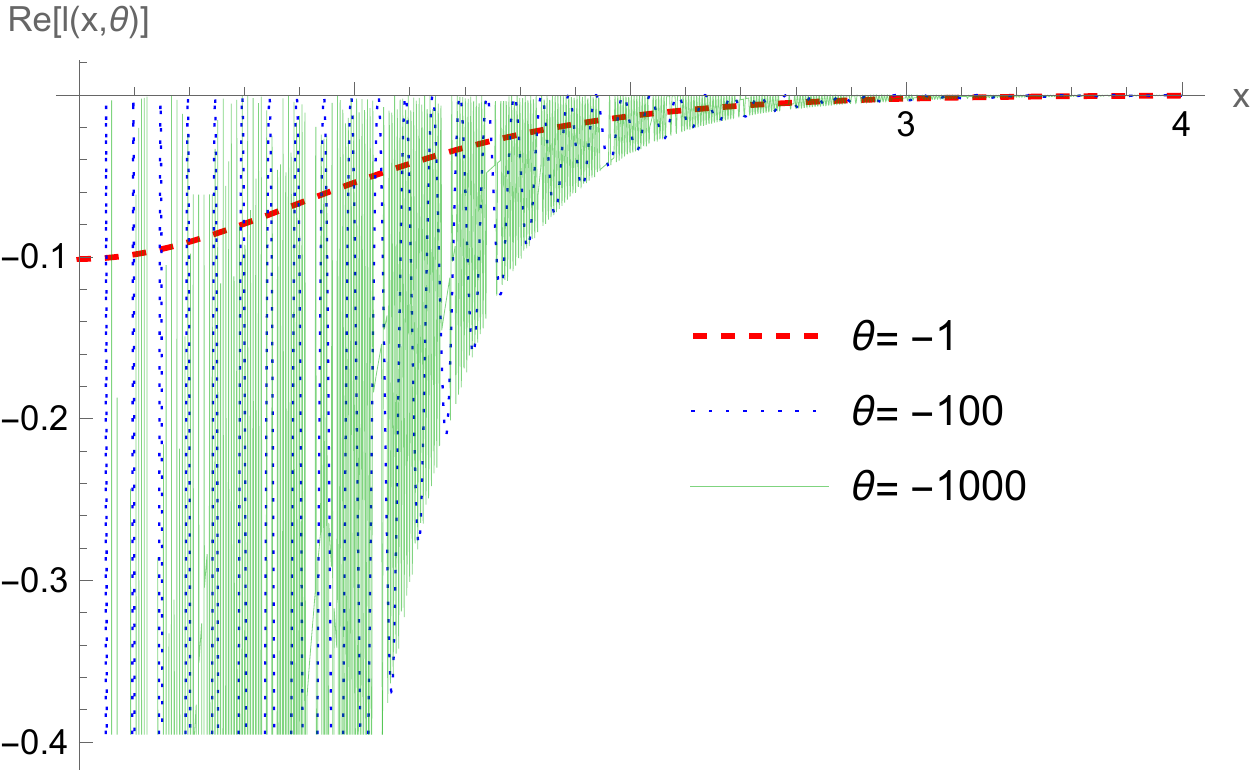}
\caption{Real part of $l(x,\theta)$ for $\theta\rightarrow -\infty$}\label{Fig:Re_l_theta-}
\end{minipage}
\end{figure}
\begin{figure}[H]
\begin{minipage}{0.48\textwidth}
\centering
\includegraphics[width=1\linewidth]{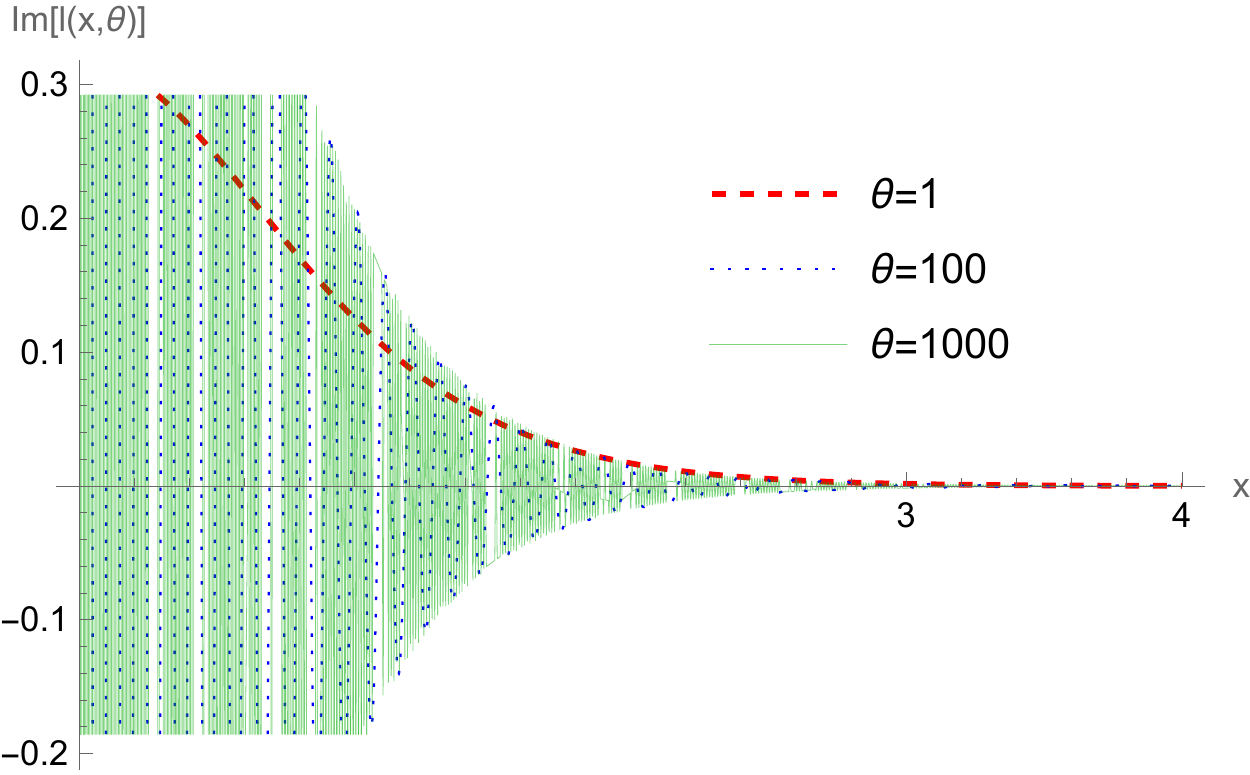}
\caption{Imaginary part of $l(x,\theta)$ for $\theta\rightarrow +\infty$}\label{Fig:Im_l_theta+}
\end{minipage}\hfill
\begin{minipage}{0.48\textwidth}
\centering
\includegraphics[width=1\linewidth]{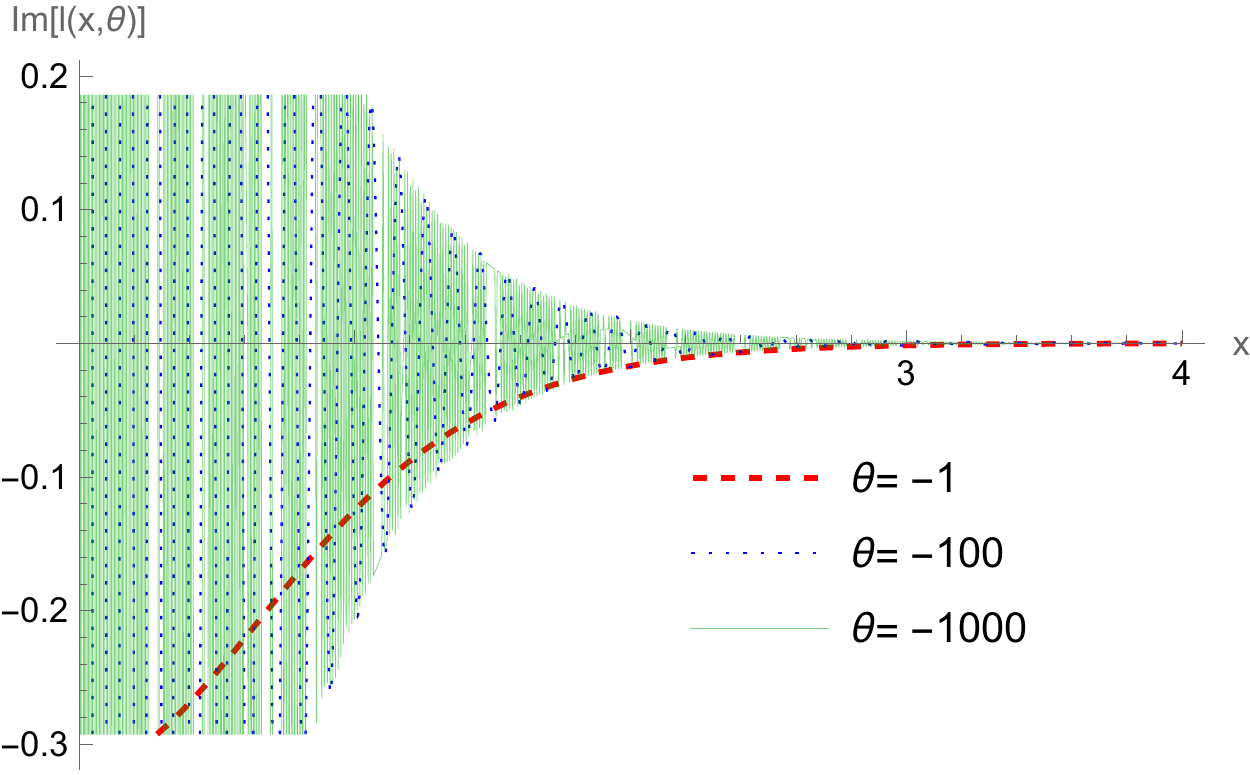}
\caption{Imaginary part of $l(x,\theta)$ for $\theta\rightarrow -\infty$}\label{Fig:Im_l_theta-}
\end{minipage}
\end{figure}
These plots highlight how both the real and imaginary parts of the integrands $f(t,\theta),l(x,\theta)$ tend to zero pretty fast in $t$ and $x$ for any value of $\theta$. For the sake of clarity we restricted the plots at $t,x=4$ but the behaviour extends on the rest of the domain of integration. For $l(x,\theta)$ we also restricted the domain shown on the $y$-axis, as both real and imaginary parts take huge values in the limit $x\rightarrow 0$. Due to the above results, in the following we shall proceed by truncating the integrals appearing in $\Phi_{DM}$ and $\Omega_{3}$ at $t,x=10$, as at this value the real and imaginary parts of the integrands $f(t,\theta),l(x,\theta)$ are already of the order $\simeq 10^{-10}$ for any $\theta$. As discussed around \eqref{Phi_expected_limits}, the plot below shows how, in the limit $\theta \rightarrow \infty$, the function $\Phi_{DM}(\theta)\rightarrow \Omega_{3}(\theta)$

\begin{figure}[H]
\begin{minipage}{0.48\textwidth}
\centering
\includegraphics[width=1\linewidth]{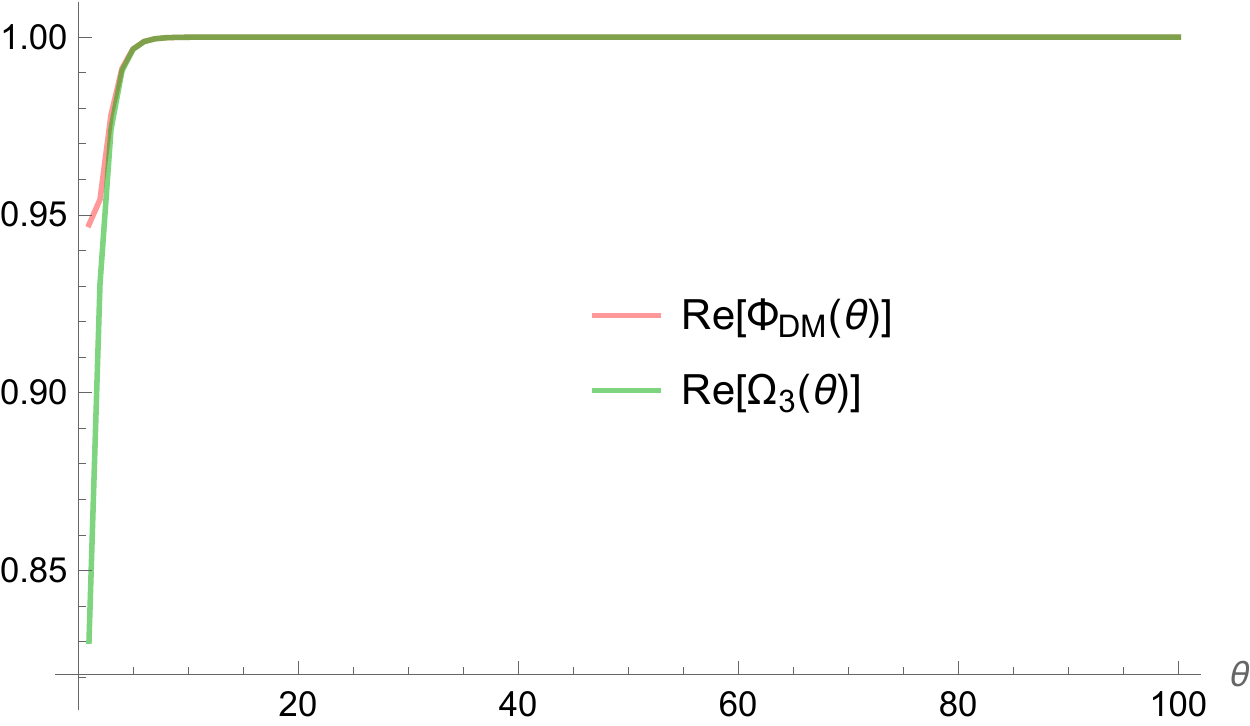}
\caption{$Re[\Phi_{DM}(\theta)]\rightarrow Re[\Omega_{3}(\theta)]$ for $\theta\rightarrow +\infty$}\label{Fig:Re_phi_to_Re_omega3}
\end{minipage}\hfill
\begin{minipage}{0.48\textwidth}
\centering
\includegraphics[width=1\linewidth]{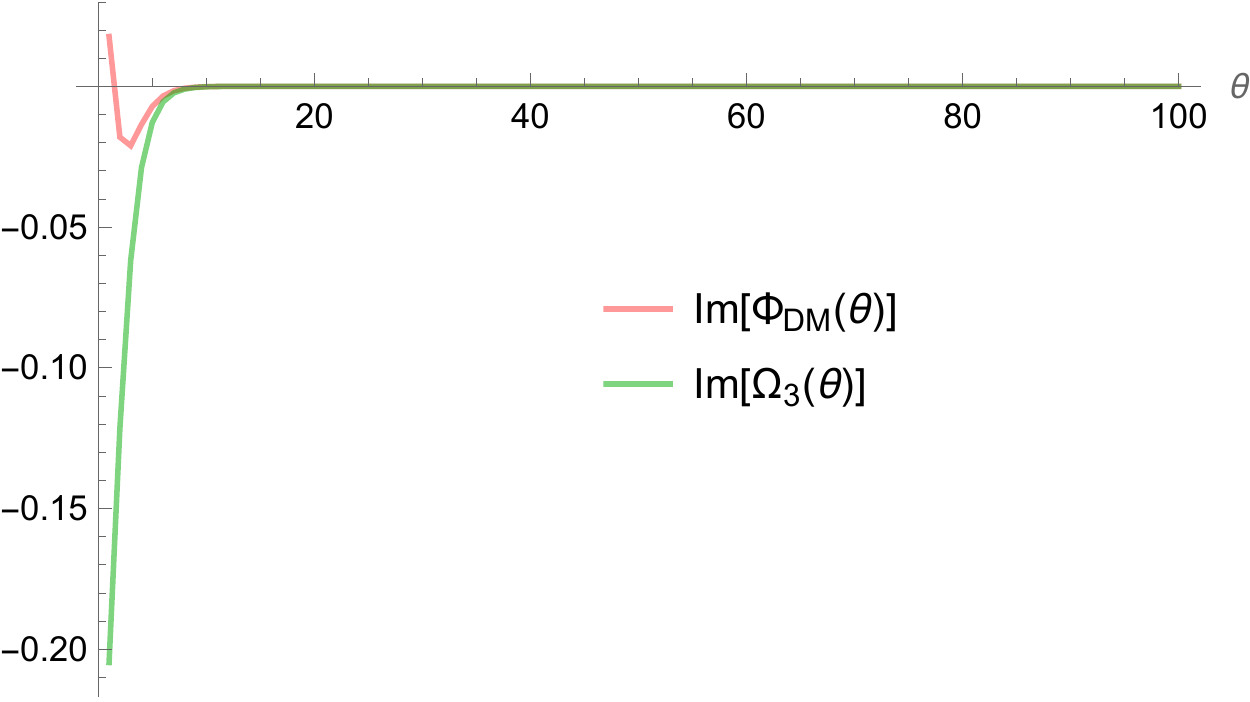}
\caption{$Im[\Phi_{DM}(\theta)]\rightarrow Im[\Omega_{3}(\theta)]$ for $\theta\rightarrow +\infty$}\label{Fig:Im_phi_to_Im_omega3}
\end{minipage}
\end{figure}

Given the limit $\Omega_{3}(\theta)\rightarrow 1$ for $\theta\rightarrow \infty$, observed in the above plots, and the relation \eqref{omega5_function_of_omega3}, one automatically has that $\Omega_{5}(\theta)\rightarrow 1$ for $\theta\rightarrow -\infty$. This implies that for $\Phi_{DM}(\theta)$ to be tending to $\Omega_{5}(\theta)$ in the limit $\theta\rightarrow -\infty$ one should find $\Phi_{DM}(\theta)\rightarrow 1$. This is indeed the case, as shown by the plots below.

\begin{figure}[H]
\begin{minipage}{0.48\textwidth}
\centering
\includegraphics[width=1\linewidth]{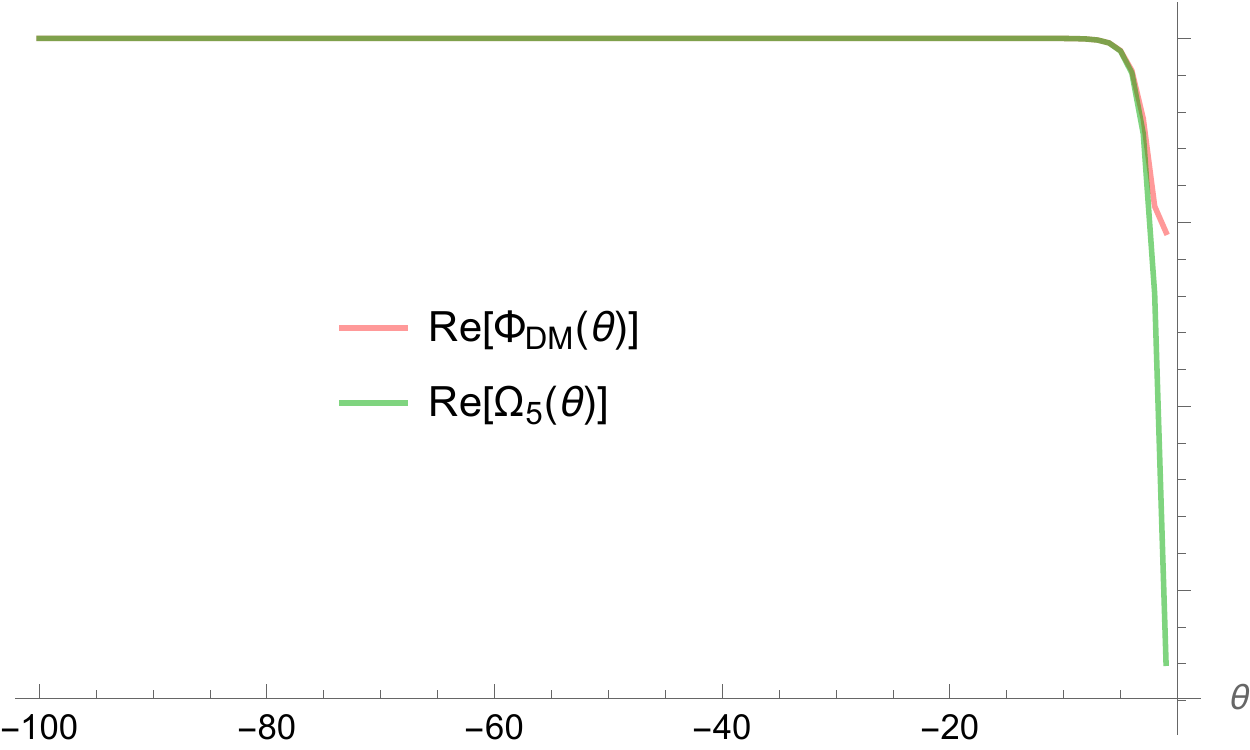}
\caption{$Re[\Phi_{DM}(\theta)]\rightarrow Re[\Omega_{5}(\theta)]$ for $\theta\rightarrow -\infty$}\label{Fig:Re_phi_to_Re_omega5}
\end{minipage}\hfill
\begin{minipage}{0.48\textwidth}
\centering
\includegraphics[width=1\linewidth]{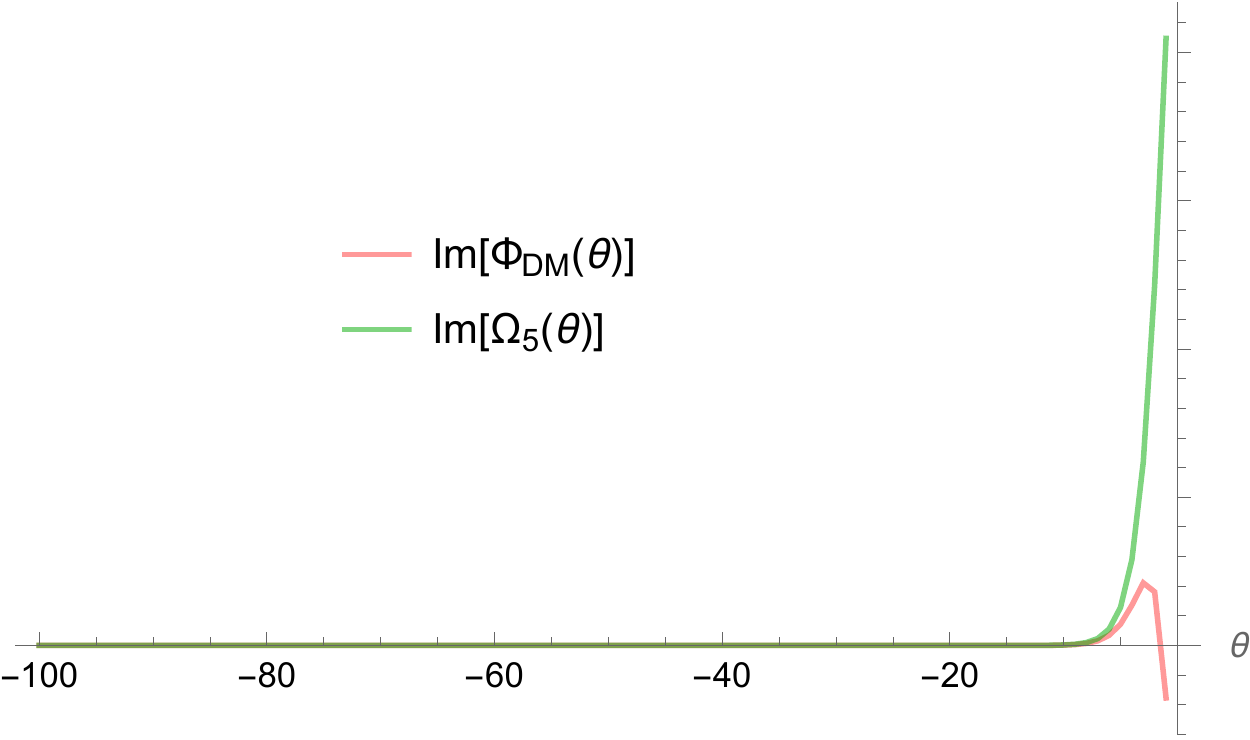}
\caption{$Im[\Phi_{DM}(\theta)]\rightarrow Im[\Omega_{5}(\theta)]$ for $\theta\rightarrow -\infty$}\label{Fig:Im_phi_to_Im_omega5}
\end{minipage}
\end{figure}

We thus conclude that the improved phase $\Phi_{DM}(\theta)$ proposed by De Martino and Moriconi in \cite{DM} does indeed tend to $\Omega_{3}(\theta)$ for $\theta\rightarrow +\infty$ and to $\Omega_{5}(\theta)$ for $\theta\rightarrow -\infty$.

\section{Conclusions}
In this paper we have initiated the study of form factors for the massless $AdS_2$ integrable scattering problem. We have focused on the purely right-right and left-left sectors to begin with, as these sectors describe a 2D CFT in the BMN limit. In the two-particle case, which is the target of our attention, the functional form of the minimal solution is expected to also extend to the non-relativistic massless $AdS_2$ scattering as well.    

We first study the ${\cal{N}}=1$ supersymmetric $S$-matrix of Fendley's \cite{Fendley:1990cy}, with the improved dressing factor found by De Martino and Moriconi \cite{DM}. We provide explicit analytic checks of the crossing and braiding/physical unitarity relations. For the first time we show numerically that the $AdS_2$ dressing factors conjectured in \cite{AA1,AA2} can be obtained as limits of the De Martino - Moriconi dressing factor.

We then follow \cite{Mussardo:1998kq} to diagonalise the $S$-matrix and propose a solution to the two-particle form factors axiom. We develop a method to obtain an integral representation of dressing factors and other quantities, which are characterised by a specific integral exponential formula inspired by \cite{a1} and involving an antisymmetric function. Our method applies quite generally, and has to possibly be understood in the distributional sense.     

As in the case of the relativistic approach to $AdS_3$ form factors \cite{Ale}, the main challenge is to find a way to test the formulas which we have obtained. Since in the BMN limit the theory is at the conformal point and presumably interactive, the scattering theory is completely non-perturbative and therefore non-perturbative methods are needed to match our proposal. In $AdS_2$, moreover, the $S$-matrix has properties that are less-standard than in $AdS_3$, and ordinary methods do not directly apply. This is why we had to resort to the Fendley $S$-matrix which is better-behaved, and argue about $AdS_2$ by taking appropriate limits.

The next step is of course to study higher-point functions. Already for ordinary ${\cal{N}}=1$ $S$-matrices, which are of $8$-vertex type, this is rendered infeasible by the absence of a pseudo-vacuum and the consequent impossibility to apply the off-shell Bethe ansatz technique of \cite{Babu,BabuF}. Once again we plan to study multi-particle form factors for the Fendley $S$-matrix in the future, using the building blocks which we have constructed in this paper. We hope that the known limits would then be well defined also for these putative expressions, which would then constitute predictions for $AdS_2$.

\begin{appendix}

\section{Proofs Of Crossing And Unitarity}\label{app:Appendix A}

\subsection{Integral method}

As a further demonstration of the technique which we have devised in section \ref{tech}, we can offer a neat way of proving the crossing and unitarity equation for $\Phi_{DM}$. This proceeds as follows using the integral representation (\ref{inteo}).

First, we can use simple trigonometric identites to combine the two exponents which appear on the respective l.h.s.s of the crossing and braiding-unitarity equation, which turn out to produce the exact same expression:

\begin{eqnarray}
\Phi_{DM}(\theta)\Phi_{DM}(\theta+i\pi)=\Phi_{DM}(\theta)\Phi_{DM}(-\theta)=\exp 2 \int_0^\infty dt \, \frac{h(t)}{t} \, \sin^2 t\theta \equiv \exp g(\theta).\label{defnt}
\end{eqnarray}
By taking a derivative we can reduce the integral to the familiar form (\ref{reca}):

\begin{eqnarray}
\frac{dg(\theta)}{d\theta}=2 \int_0^\infty dt \, h(t) \, \sin 2 t\theta = \frac{i}{\pi}\int_0^\infty dt \, h\Big(\frac{t}{2\pi}\Big) \, \sinh \frac{t \theta}{i\pi}.
\end{eqnarray}
Since $h(t)$ is an odd function, we are exactly in the situation of the assumption (\ref{simo}) which we made in order for our method to work, therefore we can simply apply the result:

\begin{eqnarray}
\frac{dg(\theta)}{d\theta} = -\sqrt{\frac{\pi}{2}} \tilde{f}\Big(\frac{\theta}{\pi}\Big), \qquad f(t) = \frac{1}{2\pi i}h\Big(\frac{t}{2\pi}\Big).
\end{eqnarray}
Mathematica knows how to do this Fourier transform:

\begin{eqnarray}
\frac{dg(\theta)}{d\theta} = \frac{1}{2}\Big(-\tanh \frac{\theta}{2} + 4 \tanh \theta -3\tanh \frac{3\theta}{2}\Big),
\end{eqnarray}
hence
\begin{eqnarray}
g(\theta) = - \log \cosh \frac{\theta}{2} + 2 \log \cosh \theta -\log \cosh \frac{3\theta}{2} + c,
\end{eqnarray}
with $c$ a constant. But this simply means that

\begin{eqnarray}
\exp g(\theta) = e^c \cosh^2 \theta \,\, \mbox{sech}\frac{\theta}{2} \,\, \mbox{sech} \frac{3\theta}{2} = \frac{1}{1+\frac{\sinh^2 \frac{\theta}{2}}{\cosh^2 \theta}},
\end{eqnarray}
where we have adjusted the constant to be $c=0$. This can be seen by the fact that $\exp g(0)$ must be equal to $1$ for the definition (\ref{defnt}) to reproduce the crossing equation. We have therefore recovered (\ref{prove1}) and (\ref{prove2}). In fact conjugating the integral expression (\ref{inteo}) is the same as changing the sign of $\theta$, and braiding and physical unitarity effectively work the same way for this dressing factor.

\subsection{Infinite product}

Let us also recap how it is possible to prove that the infinite product representation (\ref{DMM}) satisfies crossing, braiding unitarity and physical unitarity - namely (\ref{prove1}) and (\ref{prove2}). 

By simply multiplying the infinite product expression, times the same expression shifted $\theta + i \pi$, one obtains a bigger infinite product. However, a lot of the gamma functions now cancel out. The remaining gamma functions can also be made to cancel simply by repeated use of \begin{eqnarray}
\Gamma(z+1) = z \Gamma(z).\label{repeat}  
\end{eqnarray}
At the end of the massive cancellation one is left with an infinite product of the rational terms which the repeated use of
(\ref{repeat}) has left behind. Such remnants end up being exactly the same for crossing and for unitarity, and they read (assuming $\theta \in \mathbbmss{R}$)
\begin{eqnarray}
&&\Phi_{DM}(\theta)\Phi_{DM}(\theta+i \pi) =  \Phi_{DM}(\theta)\Phi_{DM}(-\theta) = \Phi_{DM}(\theta)\Phi_{DM}(\theta)^* =\cosh^2\theta\times \nonumber\\
&& \prod_{j=0}^\infty \frac{(\frac{1}{2}+j)^2(\frac{5}{2}+3j)^2(\frac{3}{2}+3j)^2(\frac{1}{2}+3j)^2}{(\frac{1}{2}+j-t)(\frac{1}{2}+j+t)(\frac{5}{2}+3j-3t)(\frac{5}{2}+3j+3t)(\frac{3}{2}+3j-3t)(\frac{3}{2}+3j+3t)(\frac{1}{2}+3j-3t)(\frac{1}{2}+3j+3t)},\nonumber
\end{eqnarray}
where we have set 
\begin{eqnarray}
t \equiv \frac{i\theta}{2\pi}    
\end{eqnarray}
(such that under crossing $t \to t - \frac{1}{2}$, while under both braiding and physical unitarity $t\to -t$).
Once again Mathematica can do this product and, once multiplied by the $\cosh^2 \theta$ in front, we obtain precisely the expression 
\begin{eqnarray}
\frac{1}{1+\frac{\sinh^2 \frac{\theta}{2}}{\cosh^2 \theta}}    
\end{eqnarray}
which we need to complete the proofs.
\end{appendix}

\section*{Acknowledgements}
AT gratefully acknowledges support from the EPSRC-SFI grant EP/S020888/1 \textit{Solving Spins and Strings}.  VG thanks STFC for Doctoral Training Programme funding (ST/W507854-2021 Maths DTP). DB was partially supported by Universit{\`a} degli studi di Milano-Bicocca, by the Italian Ministero dell'Universit{\`a} e della Ricerca (MUR), and by the Istituto Nazionale di Fisica Nucleare (INFN) through the research project `Gauge theories, Strings, Supergravity' (GSS). \\
The authors thank the anonymous referee for useful comments that have led to an improvement of the manuscript.

\section*{Data and Licence Management}
No additional research data beyond the data presented and cited in this work are needed to validate the research findings in this work. For the purpose of open access, the authors have applied a Creative Commons Attribution (CC BY) licence to any Author Accepted Manuscript version arising.

\end{document}